\documentclass[article,nojss]{jss}
\usepackage{graphicx, amsmath, amsfonts, amssymb}
\usepackage{bm} 
\usepackage{url}
\newcommand{\ARMA}{{\rm ARMA\,}}
\newcommand{\ARIMA}{{\rm ARIMA\,}}
\newcommand{\SARIMA}{{\rm SARIMA\,}}
\newcommand{\VARMA}{{\rm VARMA\,}}
\newcommand{\VARIMA}{{\rm VARIMA\,}}
\newcommand{\SVARIMA}{{\rm SVARIMA\,}}
\newcommand{\TAR}{{\rm TAR\,}}
\newcommand{\SETAR}{{\rm SETAR\,}}
\newcommand{\ARFIMA}{{\rm ARFIMA\,}}

\newcommand{\ARIMAX}{{\rm ARIMAX\,}}
\newcommand{\VMA}{{\rm VMA\,}}
\newcommand{\MA}{{\rm MA\,}}
\newcommand{\AR}{{\rm AR\,}}
\newcommand{\ARCH}{{\rm ARCH\,}}
\newcommand{\GARCH}{{\rm GARCH\,}}

\newcommand{\VEC}{{\rm ~vec\,}}
\newcommand{\BIC}{{\rm BIC\,}}
\newcommand{\AIC}{{\rm AIC\,}}

\newcommand{\FPE}{{\rm FPE\,}}
\newcommand{\HQ}{{\rm HQ\,}}
\newcommand{\SC}{{\rm SC\,}}

\newcommand{\diag}{{\rm ~diag\,}}

\newcommand{\OLS}{{\rm OLS\,}}
\newcommand{\MLE}{{\rm MLE\,}}
\newcommand{\LM}{{\rm LM\,}}
\newcommand{\GLM}{{\rm GLM\,}}
\newcommand{\US}{{\rm US\,}}
\def\VAR{\textsc{VAR}\,}

\author{Esam Mahdi\\ Qatar University}
\title{\pkg{portes}: An \proglang{R} Package for Portmanteau Tests in Time Series Models}

\Plainauthor{Esam Mahdi, Second Author} 
\Plaintitle{Portmanteau Tests for Time Series Models} 
\Shorttitle{Portmanteau Tests for Time Series Models}

\Abstract{
In this article we introduce the \proglang{R} package \pkg{portes}
\citep{MahdiMcLeodportes2015} with extensive illustrative applications.
The asymptotic distributions and the Monte Carlo procedures
of the most popular univariate and multivariate portmanteau test statistics,
including a new generalized variance statistic,
for time series models using
the powerful parallel computing framework facility 
are implemented in this package.
The proposed package has a general mechanism where user can compute the
test statistic for diagnostic checking of any time series models.
This package is also useful for simulating univariate and multivariate
seasonal and nonseasonal \ARIMA/\VARIMA time series with
finite and infinite variances, testing for stationarity and invertibility,
and estimating parameters from stable distributions.
}
\Plainkeywords{keywords, comma-separated, not capitalized, Java} 

\Address{
  Esam Mahdi\\
  Department of Mathematics, Statistics, and Physics\\
  Qatar University\\
  E-mail: \email{emahdi@qu.edu.qa}\\
}

\begin{document}


\section[Introduction]{Introduction}
\label{Introduction}

The multivariate seasonal vector integrated autoregressive moving average, \SVARIMA$\bm{(p,d,q)\times(ps,ds,qs)_s}$,
model with $k\times 1$ mean vector $\bm\mu$
and deterministic equation $\bm{a} + \bm{b}t$ for a $k$-dimensional time series
$\bm{Z}_{t}=(Z_{1,t},\ldots,Z_{k,t})^{\prime}$
can be written as
\begin{equation}\label{SVARIMA.Model1}
\bm{\Phi(B)\Phi^{\bullet}(B^s)\bigtriangledown(B)\bigtriangledown_s(B^s)}\bm{Z}_{t}-\bm\mu = \bm{a} + \bm{b} t + \bm{\Theta(B)\Theta^{\bullet}(B^s)e}_{t},
\end{equation}
where
$\bm{\Phi(B)}=\bm{\mathbb{I}}_{k}-\bm{\Phi_{1}B-\cdots-\Phi_p{B}^{p}}$,
$\bm{\Theta(B)}=\bm{\mathbb{I}}_{k}-\bm{\Theta_{1}B-\cdots-\Theta_p{B}^{q}}$,
$\bm{\Phi^{\bullet}(B^s)}=\bm {\mathbb{I}}_{k}-\bm{\Phi^{\bullet}_{1}{B^s}-\cdots-{\Phi}^{\bullet}_{ps}{B}^{ps}}$,
$\bm{\Theta^{\bullet}(B^s)}=\bm {\mathbb{I}}_{k}-\bm{\Theta^{\bullet}_{1}{B^s}-\cdots-{\Theta}^{\bullet}_{qs}{B}^{qs}}$,
and $\bm {\mathbb{I}}_{k}$ is a $k\times k$ identity matrix,
$\bm{\Phi}_\ell=(\phi_{ij,\ell})_{k\times k},~\bm{\Theta}_\ell=(\theta_{ij,\ell})_{k\times k}$,
$\bm{\Phi}^{\bullet}_{\ell s}=(\bm{\phi}^{\bullet}_{ij,{\ell s}})_{k\times k},~\bm{\Theta}^{\bullet}_{\ell s}=(\bm{\theta}^{\bullet}_{ij,{\ell s}})_{k\times k}$
are coefficient matrices, and $\bm{B}$ is a backshift operator such as
${\bm{B}}^{j}\bm{Z}_{t}=\bm{Z}_{t-j}$.
$\bm{\bigtriangledown(B)}= \diag[(1-\bm{B})^{d_{1}},\ldots,(1-\bm{B})^{d_{k}}]$ is a diagonal $k\times k$ matrix,
where $\bm{d}=(d_1,\ldots,d_k),~d_i\geq 0$ is the usual series differences whereas
$\bm{\bigtriangledown_s(B^s)}= \diag[(1-\bm{B}^s)^{ds_{1}},\ldots,(1-\bm{B}^s)^{ds_{k}}]$ is a diagonal $k\times k$ matrix,
where $\bm{ds}=(ds_1,\ldots,ds_k),~ds_i\geq 0$ is the seasonal series differences and $s$ is the seasonal period.
This states that each individual series $Z_{i}, i=1,\ldots,k$ is differenced $d_{i}\times ds_i$ times
to reduce to a stationary \VARMA$(p,q)$ series.
It is assumed that the \VARMA model is stationary,
invertible, and identifiable
\citep{Reinsel1997,Box2008}.
The $k\times 1$ coefficients $\bm{a}$ and $\bm{b}$ represent the constant drift
and the deterministic time trend respectively
and the white noise process
$\bm{e}_{t}=(e_{1,t},\ldots,e_{k,t})^{\prime}$ is assumed to be
uncorrelated in time with mean zero;
that is, $E(\bm{e}_{t})=\bm 0$
and $E(\bm{e}_{t}\bm{e}_{t-\ell}^{\prime})=\bm{\Gamma}_{0}\delta_{\ell}$,
where $\bm{\Gamma}_{0}$ is the $k\times k$ positive definite variance covariance matrix and
$\delta_{\ell}$ is the usual Kronecker delta with unity at $\ell=0$ and zero elsewhere.

In univariate time series, i.e. when $k=1$, the model in
Equation~\ref{SVARIMA.Model1} reduces to be an integrated autoregressive moving average,
\SARIMA$(p,d,q)\times(ps,ds,qs)_s$, model
\begin{equation}\label{SARIMA.Model1}
\phi(B)\Phi(B^s)\bigtriangledown(B)\bigtriangledown_s(B^s)Z_{t}-\mu= a + b t + \theta(B)\Theta(B^s)e_{t},
\end{equation}
where
$a$ and $b$, are the drift and the trend terms respectively,
$\phi(B)=1-\phi_{1}B-\cdots-\phi_{p}B^{p}, \theta(B)=1+\theta_{1}B+\cdots+\theta_{q}B^q$,
$\Phi(B^s)=1-\Phi_{1}B^s-\cdots-\Phi_{ps}B^{ps}, \Theta(B^s)=1+\Theta_{1}B^s+\cdots+\Theta_{qs}B^{qs}$,
$\bigtriangledown(B)=(1-B)^d$ is the usual differencing order, where $s$ is the seasonal period,
whereas $\bigtriangledown_s(B^s)=(1-B^s)^{ds}$ is
the seasonal differencing order, and $e_t$
is the white noise series with mean zero and variance $\sigma^2$ \citep{Hannan1969,PfaffBook2006,Box2008}.

It is important to indicate that the models in Equations \ref{SVARIMA.Model1}
and \ref{SARIMA.Model1} are always be rewritten in the Box-Jenkins
\VARIMA$\bm{(P,D,Q)}$ and \ARIMA$(P,D,Q)$ representations respectively, where
$\bm{P}= \bm{1 + p + ps\times s}, \bm{Q}= \bm{1 + q + qs\times s}$, and
$P= 1 + p + ps\times s, Q= 1 + q + qs\times s$ as follows
\begin{equation}\label{SVARIMA.Model2}
\bm{\Phi^{\sharp}(B)\bigtriangledown(B)}\bm{Z}_{t}-\bm\mu = \bm{a} + \bm{b} t + \bm{\Theta^{\sharp}(B)e}_{t},
\end{equation}
where
$\bm{\Phi^{\#}(B)}=\bm{\Phi(B)\Phi^{\bullet}(B^s)}$ and $\bm{\Theta^{\#}(B)}=\bm{\Theta(B)\Theta^{\bullet}(B^s)}$
and $\bm{\bigtriangledown(B)}=\diag[(1-\bm{B})^{d_{1}}(1-\bm{B}^s)^{ds_{1}},\ldots,(1-\bm{B})^{d_{k}}(1-\bm{B}^s)^{ds_{k}}]$,

\begin{equation}\label{SARIMA.Model2}
\phi^{\sharp}(B)\bigtriangledown(B)Z_{t}-\bm\mu = \bm{a} + \bm{b} t + \theta^{\sharp}(B)e_{t},
\end{equation}
where
$\Phi^{\#}(B)=\phi(B)\Phi(B^s)$ and $\Theta^{\#}(B)=\theta(B)\theta^{\bullet}(B^s)$
and $\bigtriangledown(B)= (1-B)^d(1-B^s)^{ds}$.

After fitting the \ARIMA or \VARIMA model using efficient estimators, the residuals, $\hat{e}_{i,t}$  where
$i=1,\ldots,k$ and $t = 1,\ldots,n$ can be obtained.
We may then use portmanteau goodness-of-fit test to test the adequacy of the fitted model by
checking whether the residuals are approximately white noise \citep{Li2004}.

The most popular portmanteau tests that were introduced by
\citet{BoxPierce1970,LjungBox1978,Hosking1980,LiMcLeod1981},
and a new portmanteau generalized
variance test based on the determinant of the standardized multivariate residual
autocorrelations \citep{PR2002,PR2006,LinMcLeod2006,MahdiMcLeod2012}
are implemented in the \proglang{R} package \pkg{portes}.
In addition, the portmanteau tests for nonlinear structure in time series
as proposed by \citet{McLeodLi1983} is also implemented in this package. 
\citet{AdlerFeldmanGallagher1998} studied \ARMA models with
identical and independent innovations from stable distribution with infinite variances and
\citet{LinMcLeod2008} developed a generalized variance portmanteau test
for such models using Monte Carlo techniques, where this test is also implemented in this package.
Moreover, this package has ability to include any test statistic, 
such as the Generalized Durbin-Watson test statistic 
for diagnostic checking of univariate linear and generalized models and 
Fisher test statistic for cyclic periodicity.
Testing for the adequacy of any fitted model, such as threshold autoregressive models, TAR, \citep{ChanRipley2012}, and
long memory from Autoregressive Fractionally Integrated Moving Average, \ARFIMA, \citep{FraleyLeischMaechlerReisenLemonte2012} is also implemented in the proposed package.
Brief descriptions for different portmanteau tests are given in Section~\ref{ComputationFunctions}.

The concern of using the portmanteau tests is that the accuracy
of the asymptotic distributions requires $n$ and $m$ large, where $n$ denotes
the length of the series and $m$ is the lag time value.
The development of powerful computers, specially those with multicore
systems, in which the \proglang{R} packages \pkg{Rmpi} \citep{YuRmpiRnews},
\pkg{snow} \citep{TierneyRossiniLiSevcikovaPackage2013}, and \pkg{parallel}
\citep{VeraJansenSuppi2008} work efficiently with parallel computing, especially with 
operating systems running new released operating systems of 
Windows version >= 8.0{\footnote{It is important to indicate that 
the operating systems Windows 8.0$+$ are already configured for multicore support. 
However, for other operating systems running on multicore processor, 
users should configure their systems in order to support the parallel computing.}}, 
has made Monte Carlo tests more affordable and recommended in such cases
as simulation experiments show that the Monte Carlo tests are usually more
accurate and powerful than asymptotic distributions
\citep{LinMcLeod2006,MahdiMcLeod2012}.
In addition, one of the advantages of the Monte Carlo significance test is that the p-value is
always exit and does not depend on the value of degrees of freedom while the p-value based on the
asymptotic chi-square distribution test is only defined for positive degrees of freedom.
The \pkg{parallel} package has been included in a direct support 
in \code{R} started with release 2.14.0 based on the work done for the
Comprehensive \proglang{R} Archive Network, \proglang{CRAN},
packages \code{multicore} and \code{snow}.

In Sections~\ref{MainFunction} and \ref{Applications} we describe the
main function, \code{portest()},
with some illustrative applications explaining the Monte Carlo
version of \citet{PR2002,MahdiMcLeod2012} portmanteau test
with implementation to \pkg{parallel} package.
This function can be used for testing whiteness of series
as well as for testing the adequacy of the fitted time series models
with finite or infinite variances involving \ARMA and \VARMA models.

To reproduce the results in these sections, one should install the \proglang{CRAN} \proglang{R}
packages \pkg{vars} version 1.5-3 \citep{PfaffStigler2013}, 
\pkg{fGarch} version 3042.83.1 \citep{Wuertz2013}, 
\pkg{FitAR} version 1.94 \citep{McLeodZhangXu2013}, 
\pkg{fracdiff} version 1.4-2 \citep{FraleyLeischMaechlerReisenLemonte2012},
\pkg{TSA} version 1.2 \citep{ChanRipley2012}, 
\pkg{forecast} version 8.7 \citep{HyndmanRazbash2015}, 
\pkg{lmtest} version 0.9-37 \citep{Hothornpackage2015}, and 
\pkg{gstat} version 2.0-2 \citep{PebesmaGraeler2015}, on
a computer configured for multicore support 
with a minimum requirement of quad-core, \code{ncores = 4}.

\begin{Schunk}
\begin{Sinput}
> ncores <- 4
\end{Sinput}
\end{Schunk}

For users who do not have configured systems for multicore processors or those who
wish to use a single core, change the previous code to \code{ncores <- 1}.

The package \pkg{portes} is available from the Comprehensive
\proglang{R} Archive Network, \proglang{CRAN}, website
at \url{http://CRAN.R-project.org/package=portes} version 4.0
and can be installed in the usual ways and is
ready to use after typing
\begin{Schunk}
\begin{Sinput}
> # install.packages("portes") is needed
> library("portes")
\end{Sinput}
\end{Schunk}
This package has been recently used in many high quality publishes refereed
papers such as \citep{GallagherFisher2015,CuiFisherWu2014,FisherGallagher2012,MahdiMcLeod2012},
with some illustrative applications.
One can use this package to simulate seasonal/nonseasonal time series from univariate \ARIMA process or
multivariate \VARIMA process with innovations of finite
or infinite variance from stable distribution as illustrated in Section \ref{Simulation}.
The simulated data may has a deterministic constant drift
and time trend term with non-zero mean.
The package \pkg{portes} may also used for testing stationarity and invertibility and for
estimating stable parameters of data from stable distribution.


\section[The main function: portest]{The main function: portest}
\label{MainFunction}

In this section we describe the main function, \code{portest()}.
This function has a general mechanism to implement any diagnostic test statistic in time series 
and linear models analysis involving the portmanteau statistics as given in
\citet{BoxPierce1970,LjungBox1978,Hosking1980,LiMcLeod1981,McLeodLi1983,
PR2002,PR2006,LinMcLeod2006,LinMcLeod2008,MahdiMcLeod2012}, and others.
The p-values can be evaluated using the Monte Carlo techniques and the approximate asymptotic
chi-square distribution.
The syntax of \code{portest()} is listed below:

\begin{CodeInput}
portest(obj,lags=seq(5,30,5),test=c("MahdiMcLeod","BoxPierce","LjungBox",
   "Hosking","LiMcLeod","other"),fn=NULL,squared.residuals=FALSE,
    MonteCarlo=TRUE,innov.dist=c("Gaussian","t","stable","bootstrap"),
     ncores=1,nrep=1000,model=list(sim.model=NULL,fit.model=NULL),
      pkg.name=NULL,set.seed=123,...)        
\end{CodeInput}


\subsection[Monte Carlo goodness-of-fit test]{Monte Carlo goodness-of-fit test}
\label{GoodnessOfFitTest}

The minimal required input of the function \code{portest()} is \code{portest(obj)}.
By selecting this argument, the generalized variance portmanteau statistic, \code{MahdiMcLeod()},
that is described in Section \ref{GeneralizedVarianceTest} will be implemented
using the Monte Carlo approach with one thousand replications on a single CPU
at default lags 5, 10, 15, 20, 25, and 30.

As an illustrative example, consider the univariate series \code{DEXCAUS}.
Data is available from the \proglang{R} Package \pkg{portes}
and refer to daily Canada/US foreign exchanges rates from
January 04, 1971 to September 05, 1996.
A complete description of the data can be retrieved simply by
typing \code{?DEXCAUS} in \proglang{R} session.
In this example, the \code{portest()} function implements the
Monte Carlo version of the statistic \code{MahdiMcLeod} with $10^3$ replications
using a single CPU and test for randomness of returns.
The results suggest that the \code{returns} series behaves like random series.
With respect to computer time, the CPU time equals to $26.61$ seconds
to get the output of this example.

\begin{Schunk}
\begin{Sinput}
> data("DEXCAUS")
> returns <- log(DEXCAUS[-1]/DEXCAUS[-length(DEXCAUS)])
> portest(returns)
\end{Sinput}
\begin{Soutput}
 lags statistic   p-value
    5  5.726436 0.2197802
   10 12.002784 0.1388611
   15 16.798015 0.1438561
   20 20.847948 0.1658342
   25 23.805231 0.2117882
   30 27.248101 0.2397602
\end{Soutput}
\end{Schunk}

For the Monte Carlo significance test, the function \code{portest()} with an
object \code{obj} of class \code{"list"} requires an additional input of two
arguments, \code{model} and \code{pkg.name}. 
The argument \code{model} is a list with two functions, \code{sim.model()} and
\code{fit.model()}.
In this case, users should write their own \proglang{R}
code of these two functions and make sure that the class output of the
\code{fit.model()} function is a \code{"list"} with at least one input argument,
\code{res}, where \code{res} stands for the residuals of the fitted model.
This list of outputs is then passes the argument of the function
\code{sim.model()}.
The output of the function \code{sim.model()} should be a
simulated univariate or multivariate time series from the fitted model resulted from
the other function \code{fit.model()}. 
The name of the \proglang{R} library involves the function used in the
fitted model should be entered via the argument \code{pkg.name}. 
For example, the \proglang{R} library \pkg{TSA} involves 
the function \code{TAR()} that we may use in 
the \code{fit.model()} function for fitting \TAR models. 
In this case, users must include the argument \code{pkg.name="TSA"}
in the \code{portest()}, where the function \code{TAR()} 
will be included in the body code of the \code{fit.model()} function.
(See the examples given in Section \ref{Applications}).

In general, the input argument \code{obj} must be an object with class
\code{"numeric"}, \code{"ts"}, \code{"matrix"}, \code{"mts \& ts"},
\code{"ar"}, \code{"arima0"}, \code{"Arima"}, \code{"ARIMA forecast_ARIMA Arima"}, 
\code{"lm"}, \code{"glm lm"}, \code{"varest"}, or \code{"list"}.
The class \code{"ar"} is associated with with a univariate and multivariate time series, whereas the other classes are used
with multivariate time series. 

The functions from the \pkg{stats} package \code{ar()} 
and \code{ar.ols()} implemented with univariate and 
multivariate time series and the functions \code{ar.burg()},
\code{ar.yw()}, and \code{ar.mle()} implemented only with 
univariate time series, return an object with a class \code{"ar"}.
The functions \code{lm()} and \code{glm()} in the \code{R} package \code{stats} produce
outputs with class \code{"lm"} and \code{"glm lm"} respectively.
The class object \code{"varest"} is associated with the output of the
multivariate function \code{VAR()} from the \proglang{R} \pkg{vars} package \citep{PfaffStigler2013}.
Note that the function \code{arima()} in the \pkg{stats} package has an
output object \code{Arima()}, whereas the function \code{arima0()}
has output object with class \code{"arima0"}, and 
these functions fit the \ARIMA models assuming that the drift term is zero. 
This may gives wrong results for \ARIMA models with nonzero intercepts, where 
such a case is a common case in finical time series.
This drawback with these functions has been fixed by 
the functions \code{Arima()} and \code{auto.arima()} in the \pkg{forecast} package
\citep{HyndmanRazbash2015} by including a new argument \code{include.drift}, where
the output of these functions has class \code{"ARIMA forecast_ARIMA Arima"}.
It is also important to indicated that, for \ARIMAX models with external exogenous variables, 
users of our package should use either \code{Arima()} 
or \code{auto.arima()} instead of \code{arima()} and \code{arima0()}.

By default, the argument \code{MonteCarlo = TRUE} in the main function,
\code{portest()}, implements
the Monte Carlo version of the portmanteau test statistic,
while the approximation asymptotic distribution
of the portmanteau test statistic will be implemented by
selecting the argument \code{MonteCarlo = FALSE}.
The argument \code{ncores = 1} will implement a single CPU, whereas
\code{ncores > 1} is used with multiple CPU's,
provided that the computer has configured system for multiple cores and the argument
\code{MonteCarlo = TRUE} is selected.

For objects with classes
\code{"ar"}, \code{"arima0"}, \code{"Arima"}, \code{"ARIMA forecast_ARIMA Arima"}, 
\code{"lm"}, \code{"gl lm"}, \code{"varest"}, and \code{"list"}
{\footnote{Note that the Generalized Durbin-Watson test statistic, is  
used with objects with classes \code{"lm"} or \code{"glm lm"}.}},
the Monte Carlo goodness-of-fit test checks the adequacy
of the fitted model under the null hypothesis
using the following steps:
\begin{description}
\item[1.] Get the residuals from the input object \code{obj} and calculate the observed
value of the portmanteau test at lag $m$, say $\mathfrak{D}_m^{0}$.
If the argument \code{squared.residuals = TRUE} is selected then the observed
value of the portmanteau test will be
calculated for the square of residuals to test
for nonlinearity structure in time series \citep{McLeodLi1983}.
\item[2.] Apply the suitable function among \code{arima.sim()}, \code{simulate.Arima()},
\code{simulate()}, and \code{varima.sim()} on the extracted estimated parameters 
from the associated \ARMA, \LM, \GLM, or \VAR models and simulate
a new time series where the exact distribution of the
extracted residuals are used for bootstrapping the simulated innovations series.
Users may write their own fitted and simulated functions
and pass these two functions via the argument \code{model=list(sim.model=NULL,fit.model=NULL)},
provided that the output of the fitted must has an
object \code{obj} with class \code{"list"} as described before.
In this step, the argument \code{innov.dist} is used to determine the distribution that 
will be used for generating the innovation of the simulated series.
The default distribution is Gaussian. 
However, users may use $t$ or stable distribution \citep{AdlerFeldmanGallagher1998,LinMcLeod2008}, 
or the nonparametric bootstrap technique.
\item[3.] Use the same function
used before to fit the same model to the simulated series.
\item[4.] Extract the residuals from this fitted model and calculate
the simulated portmanteau test
as described in step 1, say $\mathfrak{D}_m^{i}, i=1,2,\ldots$.
\item[5.] Repeat Steps 2 to 4 \code{nrep} times, where \code{nrep} is
an argument represents the number of replications needed to use in Monte Carlo test.
\end{description}
The Monte Carlo p-values associated with the different lag
values \citep[Chapter 4]{Dufour2006,DavisonHinkley1997}
are calculated using the
following formula
\begin{equation}
\label{MCequation}
       \hbox{p-value}=\frac{\#\{\mathfrak{D}_m^{i}
       \geq\mathfrak{D}_m^{0},~i=1,2,\ldots,\hbox{\code{nrep}}\}+1}{\hbox{\code{nrep}}+1}
\end{equation}

We investigate the timing for the function \code{portest()}
based on the Monte Carlo significance test for univariate and
multivariate time series of lengths $n=100,500, 1000$ 
with different lags $5, 10, 15, 20, 25, 30$.
In the univariate case, we generate \AR(1) series and then
compute the elapsed \AR(1) fitted time based on the
contributed \proglang{R} functions,
\code{FitAR()} \code{"arima0"}, \code{ar()}, \code{arima()}, \code{arima0()}, \code{Arima()},
and \code{auto.arima()} \citep{HyndmanRazbash2015}.
In the multivariate series we consider the series dimensions $k=2,3,4$
generated from the \VAR(1) process and the computed time is compared
between the fitted \VAR(1) model using the functions \code{VAR()}
and \code{ar.ols()} \citep{PfaffStigler2013}.

In addition, we compute the elapsed time of fitting 
Linear Model, \LM, using the function \code{lm()} and
the Generalized Linear Model, \GLM, using the function \code{glm()},
as well as the \GARCH models using the functions \code{garch()} 
\citep{TraplettiHornikLeBaron2015}
and \code{garchFit()} \citep{Wuertz2013}.

The times reported in Tables \ref{TimeUnivariate1}, \ref{TimeMultivariate1}, and \ref{TimeUnivariate2}
were achieved on a 2.5 GHz intel(R) Core(TM) i5 CPU running Windows 8.1.

\begin{table}[h]
\small \addtolength{\tabcolsep}{-4pt}
\begin{center}
\begin{tabular}{lcclcccccccccccc}
\noalign{\smallskip}
\noalign{\hrule}
\noalign{\smallskip}
ncores&&$n$&&\code{FitAR()}&&\code{ar()}&&\code{arima()}&&\code{arima0()}&&\code{Arima()}&&\code{auto.arima()}\\
 \noalign{\smallskip}
\noalign{\hrule}
 \noalign{\smallskip}

1&&100  && 26.93 && 24.59 && 29.11 && 29.37 && 29.53 &&  29.82\\

1&&500  && 27.21 && 24.94 && 31.16 && 31.46 && 31.61 && 31.11\\

1&&1000 && 27.96 && 25.59 && 32.80 && 32.11 && 32.16 && 31.83\\

\noalign{\smallskip}

4&&100  && 8.54 && 8.39 && 9.27 && 8.01 && 8.13 && 8.83 \\

4&&500  && 8.78 && 8.87 && 9.43 && 8.98 && 8.40 && 9.06\\

4&&1000 && 9.39 && 8.91 && 9.81 && 9.15 && 9.44 && 9.35\\

 \noalign{\smallskip}
\noalign{\hrule}
 \end{tabular}
 \caption{CPU time in seconds.
          The \proglang{R} functions
          \code{FitAR()}, \code{ar()}, \code{arima()}, \code{arima0()}, \code{Arima()},
          and \code{auto.arima()} are used to fit the \AR(1) model to
          univariate time series of lengths $n=100, 500, 1000$ and
          the \code{portest()} function is applied on the fitted model based on
          $10^3$ replications of Monte Carlo test.
          ncores denotes the number of cores are used.}
   \label{TimeUnivariate1}
 \end{center}
 \end{table}

\begin{table}[h]
\small \addtolength{\tabcolsep}{-4pt}
\begin{center}
\begin{tabular}{lcclcccccccccccc}
\noalign{\smallskip}
\noalign{\hrule}
\noalign{\smallskip}
&&&&&\multicolumn{3}{c}{\code{VAR()}}&&&\multicolumn{3}{c}{\code{ar.ols()}}\\
\noalign{\smallskip}
ncores&&$n$&&$k=2$&&$k=3$&&$k=4$&&$k=2$&&$k=3$&&$k=4$\\
 \noalign{\smallskip}
\noalign{\hrule}
 \noalign{\smallskip}
1&&100  && 52.94 && 53.58 && 54.29 && 38.39 && 38.78 && 38.16\\

1&&500  && 71.39 && 73.34 && 74.16 && 100.20 &&100.06 && 100.57\\

1&&1000 && 93.97 && 98.68 && 102.02 && 111.50 && 112.69 && 113.16\\

 \noalign{\smallskip}

4&&100  && 15.70 &&  16.34 && 16.63 &&10.65 && 10.66 && 10.82\\

4&&500  && 19.19 &&  20.75 && 22.01 &&27.76 &&28.11 && 29.70\\

4&&1000 && 25.23 &&  26.54 && 29.65 &&29.91 &&31.43 && 32.05\\

 \noalign{\smallskip}
\noalign{\hrule}
 \end{tabular}
 \caption{CPU time in seconds.
         The \proglang{R} functions
          \code{VAR()} and \code{ar.ols()} are used to fit the \VAR(1) model
          to series of lengths $n=100, 500, 1000$ with
          dimensions $k=2,3,4.$
          The \code{portest()} function is applied on the fitted model based on
          $10^3$ replications of Monte Carlo test.
          ncores denotes the number of cores are used.}
  \label{TimeMultivariate1}
 \end{center}
 \end{table}

\begin{table}[h]
\small \addtolength{\tabcolsep}{-4pt}
\begin{center}
\begin{tabular}{lcclccccccc}
\noalign{\smallskip}
\noalign{\hrule}
ncores&&$n$&&\code{garch()}&&\code{garchFit()}&&\code{lm()}&&\code{glm()}\\
 \noalign{\smallskip}
\noalign{\hrule}
 \noalign{\smallskip}

1&&100  && 68.27 && 120.39 && 13.1 && 16.82\\

1&&500  && 82.24 && 173.9 && 16.02 && 17.0\\

1&&1000 && 89.08 && 219.43 &&16.91 && 22.08\\

 \noalign{\smallskip}

4&&100  && 19.83 && 32.35 &&  4.25&&5.23\\

4&&500  && 25.35 && 42.53 &&4.93&&5.80\\

4&&1000 && 26.77 && 51.73 && 5.59 &&7.04\\

 \noalign{\smallskip}
\noalign{\hrule}
 \end{tabular}
 \caption{CPU time in seconds.
          Univariate time series of lengths $n=100, 500, 1000$ are generated from normal process
          and \proglang{R} functions
          \code{garch()} and \code{garchFit()} are used to fit the \GARCH(1,1) model. 
          The functions
          \code{lm()} and \code{glm()} are used to fit \LM and \GLM respectively.
          The \code{portest()} function is applied on the fitted model based on
          $10^3$ replications of Monte Carlo test.
          ncores denotes the number of cores are used.}
  \label{TimeUnivariate2}
 \end{center}
 \end{table}


\subsection[Monte Carlo testing for randomness]{Monte Carlo testing for randomness}
\label{RandomTest}

When the object \code{obj} has a class \code{"numeric"}, 
\code{"ts"}, \code{"matrix"}, or \code{"mts \& ts"}
provided that \code{MonteCarlo = TRUE} is selected,
the function \code{portest()} implements the Monte Carlo techniques
for testing the randomness behaviors of the object \code{obj} as follows,
\begin{description}
\item[1.] Treat the object \code{obj} as residuals and calculate
the observed value of the portmanteau as described in Step 1 before.
The argument \code{squared.residuals = TRUE} maybe selected if one
need to check for nonlinearity structure.
\item[2.] If \code{innov.dist = "stable"} then simulate an innovation
series with infinite variance from stable distribution.
If \code{innov.dist = "t"} and \code{dft}
{\footnote{dft: stands for degrees of freedom for $t$-distribution and must 
be entered as a positive integer when \code{innov.dist = "t"} is selected.}} 
is given simulate a white noise series from $t$-distribution with mean zero and estimated covariance
obtained from the object \code{obj}.
If \code{innov.dist = "bootstrap"} then the series given in the argument $obj$
will be resampled $n$ times with replacement in order to generate a nonparametric bootstrap 
innovation series, where the estimated covariance obtained from the provided object \code{obj}.
Otherwise, the default \code{innov.dist} is the Gaussian distribution that can be used to 
simulate a white noise series from a normal distribution with mean zero and estimated covariance
obtained from the object \code{obj}.
\end{description}
Step 2 will be repeated \code{nrep} times as described before.
Each time, the simulated portmanteau test is calculated for the simulated
series from step 2 and compared to the observed one.
The Monte Carlo p-values associated with the different
lag values are calculated from Equation \ref{MCequation} before.

Note that the arguments \code{lags = seq(5, 30, 5)} and the \code{set.seed = 123}
are the vector of lag auto-cross correlation coefficients used for test statistic and the recommended numbers to specify the seeds,
respectively, where users can change them to any other desirable values.
Note also that the optional arguments \code{seasonal} and \code{order} available from 
the three-dots argument \code{...} are not used with the 
Monte Carlo version of the test statistic,
but they are only used with the asymptotic distribution approach associated with specific 
object classes as we will explain in the next section.


\subsection[Asymptotic distribution significance test]{Asymptotic distribution significance test}
\label{Asymptoticdistribution}

By setting the argument \code{MonteCarlo = FALSE} in the main function,
\code{portest()}, the portmanteau test statistic based on the
asymptotic chi-distribution as we will describe in Section~\ref{ComputationFunctions}
will be implemented.

If the object \code{obj} has a class \code{"numeric"}, \code{"ts"}, \code{"matrix"}, or \code{"mts \& ts"}
then the optional argument \code{order} available from 
the three-dots argument \code{...} is needed to determine the degrees of freedom for the chi-distribution
in such a case. 
Otherwise, it will be automatically determined if the object \code{obj}
is a fitted time series model with class \code{"ar"}, \code{"arima0"}, 
\code{"Arima"}, \code{"ARIMA forecast_ARIMA Arima"}, or \code{"list"}.

In general \code{order = p + q + ps + qs}, where \code{p} and \code{q} 
are the orders of the \ARMA/\VARMA models, whereas
\code{ps} and \code{qs} are
the orders of the seasonal autoregressive and seasonal moving average respectively.
The optional argument code{seasonal} available from the three-dots argument \code{...}
is also needed to compute the seasonal portmanteau test statistic based on the
asymptotic distribution.

The argument \code{lags} is described as before
while the other arguments are not used in this case.

Note that for both types, Monte Carlo and asymptotic distribution,
when \code{squared.residuals = TRUE} is selected, then the test will apply for 
nonlinear structure in time series \citep{McLeodLi1983}.
When \code{squared.residuals = FALSE} is selected, then the test will apply on the usual residuals.

As an illustrative example, consider again the returns of the \code{DEXCAUS}
data.
We apply the \code{portest()} function using the asymptotic chi-square method,
and the results support the claim that the \code{returns} series behaves like random series.

\begin{Schunk}
\begin{Sinput}
> portest(returns, MonteCarlo = FALSE)
\end{Sinput}
\begin{Soutput}
 lags statistic        df   p-value
    5  5.726436  4.090909 0.2302146
   10 12.002784  7.857143 0.1431200
   15 16.798015 11.612903 0.1395903
   20 20.847948 15.365854 0.1566299
   25 23.805231 19.117647 0.2090410
   30 27.248101 22.868852 0.2396022
\end{Soutput}
\end{Schunk}


\subsection[Time series regression]{Time series regression}
\label{TimeSeriesRegression}

One of the fundamental assumptions in the linear regression models
is that the error terms are uncorrelated with mean zero. 
The assumption of normality is also important if one need to construct the 
confidence intervals or/and testing the hypotheses. 
In many applications, most of the regression models involve time series data that 
exhibit positive autocorrelations. 
The classical test statistic that is very useful in diagnostic checking in
time series regression and model selection
is the Durbin-Watson statistic
\citep{Durbin1950,Durbin1951,Durbin1971}.
This test statistic may be written as
\begin{equation}\label{dwtest} 
d = \displaystyle{\frac{\sum_{t=2}^{n}(\hat{e}_t-\hat{e}_{t-1})^2}{\sum_{t=1}^{n}\hat{e}_{t}^{2}}},
\end{equation}
where $\hat{e}_t, t= 1,2,\ldots,n$ are the \OLS residuals.

Under the null hypothesis of the absence of the autocorrelation of the disturbances,
in particular at lag 1, the test statistic, $d$, is a linear 
combination of chi-squared variables and 
should be close to 2, whereas small values of $d$ indicate positive correlation.

In econometric data, we have many cases in which
the error distribution is not normal with a higher-order autocorrelation than \AR(1)
or the exogenous variables are nonstochastic where the dependent variable is in a lagged
form as an independent variable.
With these cases, the Durbin-Watson test statistic using the 
asymptotic distribution is no accurate.
For such cases, we include, in our package \pkg{portes}, 
the Monte carlo version of any test statistic including the generalized Durbin-Watson test statistic.
In these cases, users need to select 
the two associated arguments \code{test = "other"} and \code{fn = NULL},
where \code{test = "other"} refers to a test statistic that is not listed before and 
\code{fn = dwt} is an associated \proglang{R} function which must 
to return the the statistic at different lags.
This function must includes at least two inputs: \code{obj} and \code{lags}, where \code{obj}
and \code{lags} are described as above.


\section[Computation portmanteau tests]{Computation portmanteau tests}
\label{ComputationFunctions}

\subsection[Box and Pierce portmanteau tests]{Box and Pierce portmanteau test}
\label{BoxPierceTest}

In the univariate time series, \citet{BoxPierce1970} introduced the portmanteau statistic
\begin{equation}\label{BoxPierceEqn}
Q_m=n\sum_{\ell=1}^{m}\Hat{r}_{\ell}^{2}
\end{equation}
where
$\Hat{r}_{\ell}=\sum_{t=\ell+1}^{n}\Hat{e}_{t}\Hat{e}_{t-\ell}/\sum_{t=1}^{n}\Hat{e}_{t}^2$,
and $\Hat{e}_{1},\ldots,\Hat{e}_{n}$ are the residuals.
This test statistic is implemented in the \proglang{R} function \code{BoxPierce()}
and can be used in the multivariate case as well.
It is approximately chi-square distribution with $k^2(m-p-q)$ degrees of freedom where $k$ represents
the dimension of the time series.
The usage of this function is extremely simple:

\begin{CodeInput}
BoxPierce(obj,lags=seq(5,30,5),order=0,season=1,squared.residuals=FALSE),
\end{CodeInput}
where the arguments of this function are as previously described.

It is important, as indicated by \citet{McLeod1978}, to use this test statistic 
for testing the seasonality with seasonal period $s$ in many applications.
The new test may obtained by replacing the lag $\ell$ in the test statistics 
given in Equation \ref{BoxPierceEqn} by $\ell s$, which is implemented in our package.   
In this case, the seasonal period $s$ is entered via the argument \code{season}, 
where \code{season = 1} is used for usual test with no seasonality check. 
Note also that \SARIMA models \code{order = p + q + ps + qs}, where \code{ps} and \code{qs} are
the orders of the seasonal autoregressive and seasonal moving average respectively.

Note that the function \code{portest()} with the
arguments \code{test = "BoxPierce", MonteCarlo = FALSE}, and \code{order = 0}
produces the same results as the function \code{BoxPierce()}.
The Monte Carlo version of this test statistic is implemented in the
function \code{portest()} as an argument \code{test = "BoxPierce"} provided
that \code{MonteCarlo = TRUE} is selected.

\subsection[Ljung and Box portmanteau tests]{Ljung and Box portmanteau test}
\label{LjungBoxTest}

\citet{LjungBox1978} modified the \citet{BoxPierce1970} test statistic as follows:
\begin{equation}\label{LjungBoxEqn}
\Hat{Q}_m=n(n+2)\sum_{\ell=1}^{m}(n-\ell)^{-1}\Hat{r}_{\ell}^{2}.
\end{equation}
This test statistic is is also asymptotically chi-square with degrees of
freedom $k^2(m-p-q)$ and implemented in the \proglang{R}
function \code{LjungBox()},

\begin{CodeInput}
LjungBox(obj,lags=seq(5,30,5),order=0,season=1,squared.residuals=FALSE),
\end{CodeInput}
where the arguments of this function are previously described.

Note that this test statistic is implemented in our package for 
testing the seasonality with seasonal period $s$ as described in Section \ref{BoxPierceTest}.

Note also that the function \code{portest()} with the arguments
\code{test = "LjungBox", MonteCarlo = FALSE}, and \code{order = 0}
produces the same results as the function \code{LjungBox()}.
The Monte Carlo version of this test statistic is implemented in the
function \code{portest()} as an argument \code{test = "LjungBox"} provided
that \code{MonteCarlo = TRUE} is selected.

In \proglang{R}, the function \code{Box.test()} was built to compute
the \citet{BoxPierce1970} and \citet{LjungBox1978} test statistics
only in the univariate case where we cannot use more than one single lag value at a time.
The proposed functions \code{BoxPierce()} and \code{LjungBox()} are more general than
\code{Box.test()} and can be used in the univariate or multivariate time series
in conjunction with a vector of different lag values; as well as, they can be
applied on output objects from time series fitted models as 
described in the vignette of the \pkg{portes} \proglang{R} package.

\subsection[Hosking portmanteau tests]{Hosking portmanteau test}
\label{HoskingTest}

\citet{Hosking1980} generalized the univariate portmanteau test statistics
given in eqns. (\ref{BoxPierceEqn}, \ref{LjungBoxEqn}) to the multivariate case.
He suggested the modified multivariate portmanteau test statistic

\begin{equation}\label{ModifiedMVTestHosking}
\Tilde{Q}_m=n^{2}\sum_{\ell=1}^{m}(n-\ell)^{-1}{\bm{\Hat{r}}_\ell}^{\prime}
({\bm{\Hat{R}}_{0}}^{-1}\otimes{\bm{\Hat{R}}_{0}}^{-1})\bm{\Hat{r}}_\ell
\end{equation}
where
$\bm{\Hat{r}}_\ell=\VEC{\bm{\Hat{R}}_\ell}^{\prime}$ is a $1\times k^2$ row
vector with rows of $\bm{\Hat{R}}_\ell$ stacked
one next to the other, and $m$ is the lag order.
The symbol $\otimes$ denotes the Kronecker product 
(\url{http://en.wikipedia.org/wiki/Kronecker_product}),
$ \bm{\Hat{R}}_\ell={\hat{\bm{L}}}^{\prime}\bm{\Hat{\Gamma}}_\ell\hat{\bm{L}}$,
$\hat{\bm{L}}{\hat{\bm{L}}}^{\prime}={\bm{\Hat{\Gamma}}_{0}}^{-1}$
where $\bm{\Hat{\Gamma}}_\ell=n^{-1}\sum_{t=\ell+1}^{n}\bm{\Hat{e}}_{t}{\bm{\Hat{e}}_{t-\ell}}^{\prime}$
is the lag $\ell$ residual autocovariance matrix.

The asymptotic distributions of $\Tilde{Q}_m$ is chi-squared with $k^2(m-p-q)$ degrees of freedom.
In \pkg{portest} package, this statistic is implemented in the function
\code{Hosking()}:

\begin{CodeInput}
Hosking(obj,lags=seq(5,30,5),order=0,season=1,squared.residuals=FALSE),
\end{CodeInput}
where the arguments of this function is described as before.
Note that the function \code{portest()} with the arguments
\code{test = "Hosking", MonteCarlo = FALSE}, and \code{order = 0}
produces the same results of the function \code{Hosking()}.
The Monte Carlo version of this test statistic is implemented in the
function \code{portest()} as an argument \code{test = "Hosking"} provided
that \code{MonteCarlo = TRUE} is selected.

\subsection[Li and McLeod portmanteau tests]{Li and McLeod portmanteau test}
\label{LiMcLeodTest}

\citet{LiMcLeod1981} suggested the multivariate modified portmanteau test statistic

\begin{equation}\label{ModifiedMVTestLiMcLeod}
\Tilde{Q}^{(L)}_{m}=n\sum_{\ell=1}^{m}{\bm{\Hat{r}}_\ell}^{\prime}
                   ({\bm{\Hat{R}}_{0}}^{-1}\otimes{\bm{\Hat{R}}_{0}}^{-1})
                   \bm{\Hat{r}}_\ell+ \frac{k^{2}m(m+1)}{2n}
\end{equation}

which is asymptotically distributed as chi-squared with $k^2(m-p-q)$ degrees of freedom.
In \pkg{portes} package, the test statistic $\Tilde{Q}^{(L)}_{m}$ is
implemented in the function \code{LiMcLeod()},

\begin{CodeInput}
LiMcLeod(obj,lags=seq(5,30,5),order=0,season=1,squared.residuals=FALSE),
\end{CodeInput}
where the arguments of this function is described as before.

Note that the function \code{portest()} with the arguments
\code{test = "LiMcLeod", MonteCarlo = FALSE}, and \code{order = 0}
produces the same results as the function \code{LiMcLeod()}.
The Monte Carlo version of this test statistic is implemented in the
function \code{portest()} as an argument \code{test = "LiMcLeod"} provided
that \code{MonteCarlo = TRUE} is selected.

\subsection[Generalized variance portmanteau test]{Generalized variance portmanteau test}
\label{GeneralizedVarianceTest}

\citet{PR2002} proposed a univariate portmanteau test of goodness-of-fit
test based on the $m$-th root of the
determinant of the Toeplitz residual autocorrelation matrix of order $m+1$,

\begin{equation}\label{PenaRodrigueztoeplitz}
      \mathcal{\hat{R}}_{\mathit{m}}=\left(%
\begin{array}{cccc}
  \hat{r}_{0} & \hat{r}_{1} & \ldots & \hat{r}_{m} \\
  \hat{r}_{-1} & \hat{r}_{0} & \ldots &\hat{r}_{m-1} \\
  \vdots & \ldots & \ddots &  \vdots  \\
  \hat{r}_{-m}  & \hat{r}_{-m+1} & \dots & \hat{r}_{0} \\
\end{array}%
\right)
\end{equation}
where $\hat{r}_{0}=1$ and $\hat{r}_{-\ell}=\hat{r}_{\ell},$ for all $\ell$.
They approximated the distribution of their proposed test statistic by the gamma distribution
and provided simulation experiments to demonstrate the improvement
of their statistic in comparison with the one that is given in Equation \ref{LjungBoxEqn}.

\citet{PR2006} suggested to modify the generalized variance test by taking the log of
the $(m+1)$-th root of the determinant of $\mathcal{\hat{R}}_{\mathit{m}}$ given in
Equation \ref{PenaRodrigueztoeplitz}.
They proposed two approximations by using the Gamma and Normal distributions to
the asymptotic distribution of this test and
indicated that the performance of both approximations for checking the
goodness-of-fit in linear models is similar and more powerful
for small sample size than the previous one.

\citet{LinMcLeod2006} introduced the Monte Carlo version of \citet{PR2002} and \citet{PR2006} tests
as they noted that it is quite often that
the generalized variance portmanteau test does not
agree with the suggested Gamma approximation.
They show that the Monte Carlo is more powerful than its
competitors with the correct size level.

\citet{MahdiMcLeod2012} extended these tests
to the multivariate time series and their
simulation experiments illustrated and demonstrated the usefulness
of the Monte Carlo test as well as its improved power performance compared
to the previously used multivariate portmanteau diagnostic checks.

The new multivariate generalized portmanteau test statistic proposed by \citet{MahdiMcLeod2012} is
\begin{equation}
\label{NewTest}
\mathfrak{D}_m=-3n(2m+1)^{-1}\log\mid \bm{\mathfrak{\hat{R}}}_{\mathit{m}}\mid
\end{equation}
where
\begin{equation}\label{MahdiMcLeodtoeplitz1}
      \hat{\mathfrak{\bm{R}}}_\mathit{m} = \left(%
\begin{array}{cccc}
  \mathbb{I}_k & \hat{\bm{R}}_{1} & \ldots & \hat{\bm{R}}_{\mathit{m}} \\
  \hat{\bm{R}}^{\prime}_{1} & \mathbb{I}_k & \ldots &\hat{\bm{R}}_{\mathit{m}-1} \\
  \vdots & \ldots & \ddots &  \vdots  \\
  \hat{\bm{R}}^{\prime}_{\mathit{m}} & \hat{\bm{R}}^{\prime}_{\mathit{m}-1}& \dots & \mathbb{I}_k \\
\end{array}%
\right),
\end{equation}

Replacing $\mathfrak{\Hat{\bm{R}}}_\mathit{m}$ that is given in Equation \ref{MahdiMcLeodtoeplitz1} by
$\mathfrak{\Hat{\bm{R}}}_\mathit{m}(s)$ will easily extend to test for seasonality with period $s$,
where 
\begin{equation}\label{MahdiMcLeodtoeplitz2}
      \mathfrak{\Hat{\bm{R}}}_\mathit{m}(s) = \left(%
\begin{array}{ccccc}
  \mathbb{I}_k & \Hat{\bm{R}}_{s} & \Hat{\bm{R}}_{2s} &\ldots & \bm{\Hat{R}}_{\mathit{ms}} \\
  \Hat{\bm{R}}^{\prime}_{s} & \mathbb{I}_k & \Hat{\bm{R}}^{\prime}_{s}& \ldots &\Hat{\bm{R}}_{(\mathit{m}-1)s} \\
  \vdots & \ldots & \ddots &  \ddots &\vdots  \\
  \Hat{\bm{R}}^{\prime}_{\mathit{ms}} & \Hat{\bm{R}}^{\prime}_{(\mathit{m}-1)s}& \Hat{\bm{R}}_{(\mathit{m}-2)s} &\dots & \mathbb{I}_k \\
\end{array}%
\right)
\end{equation}

The null distribution is approximately
$\chi^{2}$ with ${k^2(1.5m(m+1)(2m+1)^{-1}-o)}$ degrees of freedom
where $o= p+q+ps+qs$ denotes the order of the series as described before.
This test statistics
is implemented in the contributed \proglang{R} function \code{MahdiMcLeod()},

\begin{CodeInput}
MahdiMcLeod(obj,lags=seq(5,30,5),order=0,season=1,squared.residuals=FALSE),
\end{CodeInput}
where the arguments of this function are described as before.

Note that the function \code{portest()} with the arguments
\code{test = "MahdiMcLeod", MonteCarlo = FALSE}, and \code{order = 0}
produces the same results as the function \code{MahdiMcLeod()}.
The Monte Carlo version of this test statistic is implemented in the
function \code{portest()} as an argument \code{test = "MahdiMcLeod"} provided
that \code{MonteCarlo = TRUE} is selected.

Figure \ref{siglevel} below illustrates the accuracy of using the
Monte Carlo significance test of $\mathfrak{D}_m$ for simulated Gaussian bivariate \VAR(1) process
$\bm{Z}_{t}=\bm{\Phi}\bm{Z}_{t-1}+\bm{a}_{t}$ with mean zero and covariance matrix,
$
\bm{\Gamma}_{0}=\left(%
\begin{array}{cc}
  1 & x \\
  x & 1 \\
\end{array}%
\right)$,
where $x=0.25,0.5,0.75$.
The coefficient matrix is
$
\bm{\Phi}=\left(%
\begin{array}{cc}
  0.3 & 0.5 \\
  0 & 0.3 \\
\end{array}%
\right)
$. 
More simulation results including the simulation \proglang{R} code that we
used to compare the asymptotic distribution with the Monte Carlo significance tests are
available from the online vignette documentation of the \pkg{portes} \proglang{R} package.

\begin{figure}[h]
\begin{center}
\includegraphics{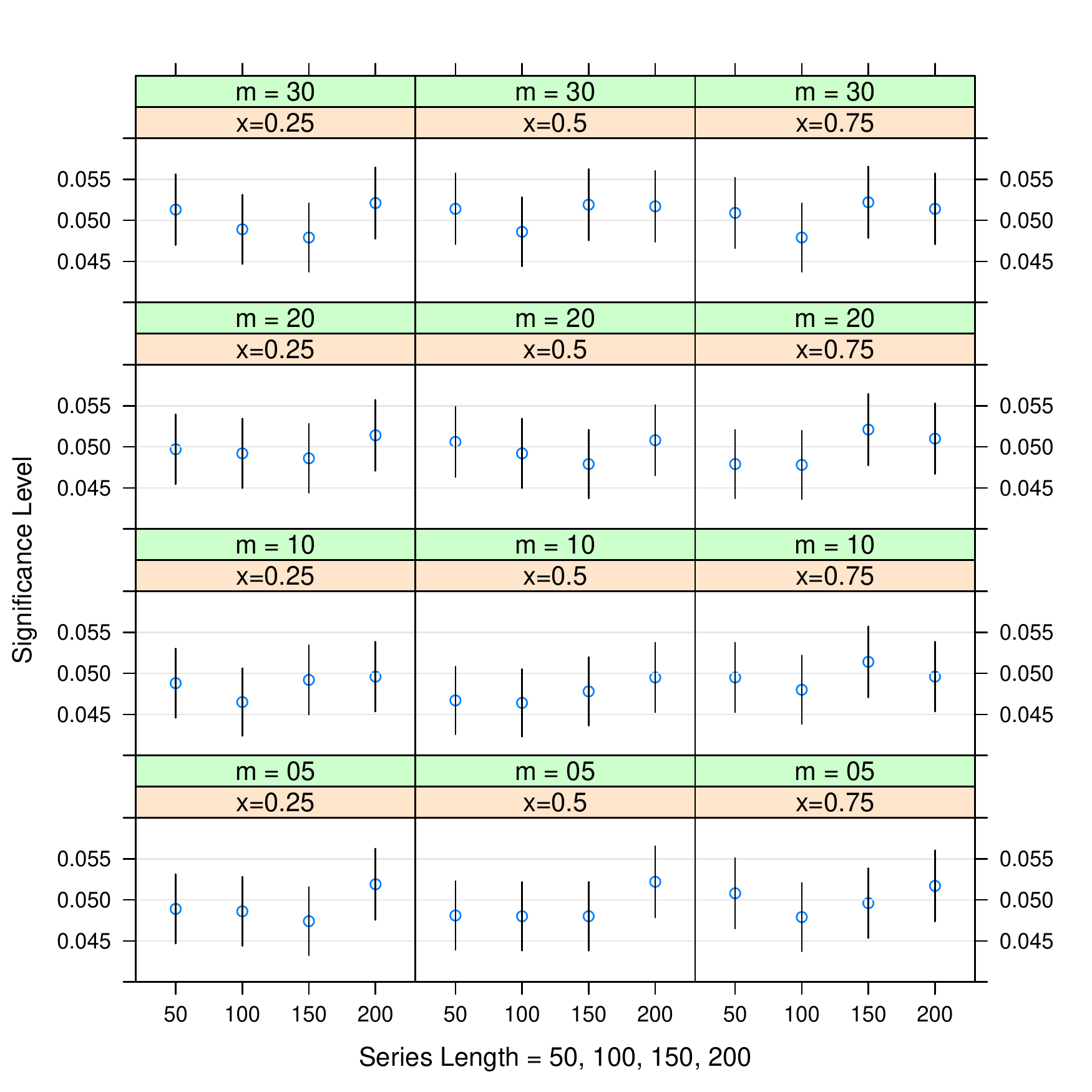}
\end{center}
\caption{The $95\%$ confidence interval of the $5\%$ level based on 
the $\mathfrak{D}_m$ statistic with $10^3$ iterations of $10^4$ simulations. 
The empirical significance level was estimated for different lag 
values, $m=5,10,20,30,$ and series lengths, $n=50,100,150,200$.}
\label{siglevel}
\end{figure}


\section[Applications]{Applications}
\label{Applications}

In this section we apply the main function \code{portest()} on some 
real data using the approximation asymptotic distribution and
the Monte Carlo significance test.
For the Monte Carlo significance test we use
the argument \code{ncores = 4} on a personal computer with 4 CPUs running on
2.5 GHz intel(R) Core(TM) i5 CPU with operating system Windows 8.1.
Users may change this number to meet the number of CPUs that they want.
In this section, we test the adequacy of the fitted models 
with output classes \code{"varest"}, \code{"list"}, \code{"lm"}, and \code{"ARIMA forecast_ARIMA Arima"}.
Testing for other classes such as \code{"numeric"}, \code{"ts"}, 
\code{"matrix"}, \code{"mts \& ts"}, \code{"ar"}, \code{"arima0"}, and \code{"Arima"} 
is straightforward and users may consult the extensive applications that are available from the online 
vignette documentation of the \pkg{portes} \proglang{R} package.


\subsection[Monte Carlo significance test for VAR models]{Monte Carlo significance test for \VAR models}
\label{Example1}

The first application make use of the multivariate macro economic data set for Canada.
Data starts at the first quarter of 1980 and ends at the fourth quarter of 2000 and available
from the statistical software \pkg{JMulTi} \citep{BreitungBruggemannLutkepohl2004} and the
\proglang{R} Package \pkg{vars} \citep{PfaffStigler2013}.

\begin{Schunk}
\begin{Sinput}
> # install.packages("vars") is needed
> require("vars")
> data("Canada")
\end{Sinput}
\end{Schunk}

According to \citet{BreitungBruggemannLutkepohl2004} and \citet{PfaffPaper2008},
the \AIC and \FPE determine the optimal lag number of an unrestricted \VAR model with
a maximal lag length of eight to be $p = 3$, whereas the \HQ criterion
indicates $p = 2$ and the \SC criterion indicates an optimal lag length of $p = 1$.
After that, they estimated for all three lag orders a \VAR including
a drift constant and a deterministic trend
and conducted diagnostic tests using the multivariate version of
the \code{BoxPierce} portmanteau test statistic.
The \code{BoxPierce} statistic suggests an adequacy in the
fitted \VAR(1) model at lag $m=12$ and 16,
it shows inadequacy at lower lags.
See results in \citep[Table 3]{PfaffPaper2008}.

In this example, instead of using \code{BoxPierce} statistic, 
we use the \code{MahdiMcLeod} portmanteau test statistic 
based on the asymptotic distribution and the Monte Carlo significance test, 
which is clearly indicates that the 
fitted \VAR(1) model is an inadequate model
 
\begin{Schunk}
\begin{Sinput}
> p1ct <- vars::VAR(Canada, p = 1, type = "both")
> portest(p1ct, lags = seq(4, 16, 4), MonteCarlo=FALSE)
\end{Sinput}
\begin{Soutput}
 lags statistic        df      p-value
    4  134.1742  37.33333 7.578382e-13
    8  191.9122  85.64706 4.040268e-10
   12  249.2777 133.76000 5.412381e-09
   16  310.9415 181.81818 7.971696e-09
\end{Soutput}
\begin{Sinput}
> portest(p1ct, test= "MahdiMcLeod", lags = seq(4, 16, 4), ncores = ncores)
\end{Sinput}
\begin{Soutput}
 lags statistic     p-value
    4  134.1742 0.000999001
    8  191.9122 0.000999001
   12  249.2777 0.002997003
   16  310.9415 0.020979021
\end{Soutput}
\end{Schunk}

Next, we fit a \VAR(3) with a constant and a trend using the
function \code{VAR} and implement the goodness of fit tests
based on the Monte Carlo version of \code{MahdiMcLeod} statistic,
and the result suggests that the fitted \VAR(3) model is an adequate model.

\begin{Schunk}
\begin{Sinput}
> p3ct <- vars::VAR(Canada, p = 3, type = "both")
> portest(p3ct, lags = seq(4, 16, 4), ncores = ncores)
\end{Sinput}
\begin{Soutput}
 lags statistic   p-value
    4  19.70053 0.8431568
    8  57.65003 0.9670330
   12 109.50000 0.9860140
   16 175.40899 0.9940060
\end{Soutput}
\end{Schunk}

It is important to indicate that we implement the 
\code{MahdiMcLeod} portmanteau test statistic on the fitted model 
in order to get efficient and reliable results. 
However, users can always implement this test on the 
fitted residual resulted from the fitted models such 
as \code{p1ct} and \code{p3ct} as follows

\begin{Schunk}
\begin{Sinput}
> res1 <- resid(p1ct) 
> portest(res1, lags = seq(4, 16, 4), MonteCarlo=FALSE, order = 1)
\end{Sinput}
\begin{Soutput}
 lags statistic        df      p-value
    4  134.1742  37.33333 7.578382e-13
    8  191.9122  85.64706 4.040268e-10
   12  249.2777 133.76000 5.412381e-09
   16  310.9415 181.81818 7.971696e-09
\end{Soutput}
\begin{Sinput}
> portest(res1, lags = seq(4, 16, 4), ncores = ncores)
\end{Sinput}
\begin{Soutput}
 lags statistic     p-value
    4  134.1742 0.000999001
    8  191.9122 0.000999001
   12  249.2777 0.001998002
   16  310.9415 0.024975025
\end{Soutput}
\begin{Sinput}
> res3 <- resid(p3ct)
> portest(res3, lags = seq(4, 16, 4), ncores = ncores)
\end{Sinput}
\begin{Soutput}
 lags statistic p-value
    4  19.70053       1
    8  57.65003       1
   12 109.50000       1
   16 175.40899       1
\end{Soutput}
\begin{Sinput}
> detach(package:vars)
\end{Sinput}
\end{Schunk}


\subsection[Monte Carlo significance test for subset autoregression models]{Monte Carlo significance test for subset autoregression models}
\label{Example2}

This application uses the univariate series of the logarithms of
Canadian lynx trappings from 1821 to 1934.
Data is available from the \proglang{R} package \pkg{datasets} under the name \code{lynx}.
The \pkg{FitAR} package provides an efficient and reliable exact \MLE for 
estimating the \AR and the subset \AR models for such a data
\citep{McLeodZhangXu2013,McLeodZhangPaper2008}, where 
the \AR(2) fitted model was selected based on the \BIC criterion implemented from
the \code{SelectModel()} function available from the \pkg{FitAR} package.

\begin{Schunk}
\begin{Sinput}
> # install.packages("FitAR") is needed
> require("FitAR")
> lynxData <- log(lynx)
>  p <- SelectModel(lynxData, ARModel = "AR", Criterion = "BIC",Best = 1)
>   fit <- FitAR(lynxData, p, ARModel = "AR")
\end{Sinput}
\end{Schunk}

However, applying the Monte Carlo procedures on the fitted model are more accurate and preferable 
than applying these procedures on the fitted residual, but to get fast results users
can apply either Monte Carlo or asymptotic distribution portmanteau tests on residuals as follows 

\begin{Schunk}
\begin{Sinput}
>   res <- fit$res
>  MahdiMcLeod(res, order = p) 
\end{Sinput}
\begin{Soutput}
 lags statistic        df     p-value
    5  5.984989  2.090909 0.054687987
   10 10.036630  5.857143 0.115222212
   15 21.447021  9.612903 0.014964682
   20 31.810564 13.365854 0.003100578
   25 38.761595 17.117647 0.002040281
   30 43.936953 20.868852 0.002252062
\end{Soutput}
\begin{Sinput}
>  portest(res,MonteCarlo = FALSE, order = p)
\end{Sinput}
\begin{Soutput}
 lags statistic        df     p-value
    5  5.984989  2.090909 0.054687987
   10 10.036630  5.857143 0.115222212
   15 21.447021  9.612903 0.014964682
   20 31.810564 13.365854 0.003100578
   25 38.761595 17.117647 0.002040281
   30 43.936953 20.868852 0.002252062
\end{Soutput}
\begin{Sinput}
>  portest(res,ncores = ncores)
\end{Sinput}
\begin{Soutput}
 lags statistic     p-value
    5  5.984989 0.194805195
   10 10.036630 0.238761239
   15 21.447021 0.033966034
   20 31.810564 0.008991009
   25 38.761595 0.002997003
   30 43.936953 0.001998002
\end{Soutput}
\end{Schunk}
The output suggest that we should reject the null hypothesis of the absence of the autocorrelation
and the model is not good.

Because the \code{portest()} function does not directly support 
the fitted models with class \code{"FitAR"} obtained by the \code{FitAR()} 
function from the \pkg{FitAR} package, users must coded two functions as described before in
Section \ref{GoodnessOfFitTest}.
One for fitting the model with output object of class \code{"list"}
with at least one components, \code{res}, where \code{res} stands for the residuals of the fitted model. 
This function feeds in the argument of the other function that coded for performing 
the Monte Carlo simulation.
In this regard, we introduce the following code of two functions as an example 
for testing the autocorrelation in fitted residual based
on the Monte Carlo version of the portmanteau $\mathfrak{D}_m$ statistic 
using the \code{portest()} function.
    
\begin{Schunk}
\begin{Sinput}
> FitARModel <- function(data){
+     p <- SelectModel(data, ARModel = "AR", Criterion = "BIC",Best = 1)
+     fit <- FitAR(data, p, ARModel = "AR")
+     res <- fit$res
+     phiHat <- fit$phiHat
+     sigsqHat <- fit$sigsqHat
+  list(res=res,order=p,phiHat=phiHat,sigsqHat=sigsqHat)
+ }
> SimARModel <- function(model){
+           phi <- model$phiHat
+           n <- length(model$res)
+           sigma <- model$sigsqHat
+        ts(SimulateGaussianAR(phi, n = n, InnovationVariance = sigma))
+ }
\end{Sinput}
\end{Schunk}

The function \code{SimARModel()} is the one that users need to 
simulate from fitted \AR models obtained from the other function \code{FitARModel()}
in order to calculate the p-values based on the Monte Carlo significance test.
Note that the function \code{FitARModel()} may used to calculate the p-values based on the asymptotic distribution,
but the other one, \code{SimARModel()}, is not needed for such a case and only needed for the Monte Carlo significance test.  

The asymptotic distribution and the Monte Carlo version of $\mathfrak{D}_m$ test 
are then applied as follows 

\begin{Schunk}
\begin{Sinput}
> Fit <- FitARModel(lynxData)
> portest(Fit,MonteCarlo=FALSE) ## Asymptotic distribution of MahdiMcLeod test
\end{Sinput}
\begin{Soutput}
 lags statistic        df     p-value
    5  5.984989  2.090909 0.054687987
   10 10.036630  5.857143 0.115222212
   15 21.447021  9.612903 0.014964682
   20 31.810564 13.365854 0.003100578
   25 38.761595 17.117647 0.002040281
   30 43.936953 20.868852 0.002252062
\end{Soutput}
\begin{Sinput}
> portest(Fit,test = "MahdiMcLeod", ncores = ncores,
+   model=list(sim.model=SimARModel,fit.model=FitARModel),pkg.name="FitAR")
\end{Sinput}
\begin{Soutput}
 lags statistic     p-value
    5  5.984989 0.033966034
   10 10.036630 0.059940060
   15 21.447021 0.005994006
   20 31.810564 0.000999001
   25 38.761595 0.000999001
   30 43.936953 0.000999001
\end{Soutput}
\end{Schunk}

For lags $m \geq 15$, the $\mathfrak{D}_m$ statistic based on 
both approaches suggests model inadequacy.
Fitting a subset autoregressive using the function \code{FitARp()}
with the \BIC criterion,
the Monte Carlo version of the portmanteau test 
more clearly suggests model adequacy, whereas the asymptotic chi-square
suggests inadequacy at lags $15$ and adequacy otherwise.

\begin{Schunk}
\begin{Sinput}
> FitsubsetAR <- function(data){
+     FitsubsetAR <- FitARp(data, c(1, 2, 4, 10, 11))
+     res <- FitsubsetAR$res
+     phiHat <- FitsubsetAR$phiHat
+     p <- length(phiHat)
+     sigsqHat <- FitsubsetAR$sigsqHat
+  list(res=res,order=p,phiHat=phiHat,sigsqHat=sigsqHat)
+ }
> SimsubsetARModel <- function(parSpec){
+           phi <- parSpec$phiHat
+           n <- length(parSpec$res)
+           sigma <- parSpec$sigsqHat
+        ts(SimulateGaussianAR(phi, n = n, InnovationVariance = sigma))
+ }
> Fitsubset <- FitsubsetAR(lynxData)
> portest(Fitsubset,MonteCarlo=FALSE)
\end{Sinput}
\begin{Soutput}
 lags statistic         df     p-value
    5  2.374225  0.0000000          NA
   10  3.598248  0.0000000          NA
   15  5.661285  0.6129032 0.008190694
   20  8.590962  4.3658537 0.090004731
   25 11.462473  8.1176471 0.184353957
   30 13.900470 11.8688525 0.297764350
\end{Soutput}
\begin{Sinput}
> portest(Fitsubset,test = "MahdiMcLeod", ncores = ncores,
+   model=list(sim.model=SimsubsetARModel,fit.model=FitsubsetAR),pkg.name="FitAR")
\end{Sinput}
\begin{Soutput}
 lags statistic   p-value
    5  2.374225 0.3846154
   10  3.598248 0.6923077
   15  5.661285 0.7422577
   20  8.590962 0.7112887
   25 11.462473 0.6853147
   30 13.900470 0.7042957
\end{Soutput}
\begin{Sinput}
> detach(package:FitAR)
\end{Sinput}
\end{Schunk}


\subsection[Monte Carlo significance test for ARCH effects]{Monte Carlo significance test for \ARCH effects}
\label{Example3}

Our next application is given as an illustrative application of
testing for Autoregressive Conditional Heteroscedastic, \ARCH, effects.
We consider the monthly log stock returns of Intel Corporation
dataset from January 1973 to December 2003.
Data has been discussed by \citet[p.99-102]{Tsay2005},  
which is available from the package \pkg{portes} with the names \code{monthintel}.
First we apply the asymptotic distribution and the Monte Carlo version of the statistic
\code{MahdiMcLeod} directly on the returns by considering the 
argument \code{squared.residuals=FALSE} of the \code{portest()} function, which
suggests no significant serial correlations.

\begin{Schunk}
\begin{Sinput}
> data("monthintel")
> monthintel <- as.ts(monthintel)
> portest(monthintel,lags=seq(10,40,10),MonteCarlo=FALSE,squared.residuals=FALSE)
\end{Sinput}
\begin{Soutput}
 lags statistic        df   p-value
   10   9.85822  7.857143 0.2632960
   20  20.38081 15.365854 0.1738108
   30  28.17267 22.868852 0.2039194
   40  34.86373 30.370370 0.2627128
\end{Soutput}
\begin{Sinput}
> portest(monthintel, lags = seq(10, 40, 10), MonteCarlo = TRUE, 
+         ncores = ncores, squared.residuals = FALSE)
\end{Sinput}
\begin{Soutput}
 lags statistic   p-value
   10   9.85822 0.2437562
   20  20.38081 0.1518482
   30  28.17267 0.1668332
   40  34.86373 0.2057942
\end{Soutput}
\end{Schunk}

After that we apply both methods of \code{MahdiMcLeod} on the squared returns 
by considering the argument \code{squared.residuals=TRUE} of the \code{portest()} function.
The result suggests that the monthly returns are not serially uncorrelated, but the return series
are dependent due to existence of an Autoregressive Conditional Heteroscedastic, \ARCH, effects

\begin{Schunk}
\begin{Sinput}
> portest(monthintel,lags=seq(10,40,10),MonteCarlo=FALSE,squared.residuals=TRUE)
\end{Sinput}
\begin{Soutput}
 lags statistic        df      p-value
   10  34.67590  7.857143 2.718240e-05
   20  52.04751 15.365854 7.097281e-06
   30  64.65689 22.868852 7.331587e-06
   40  72.26830 30.370370 2.856747e-05
\end{Soutput}
\begin{Sinput}
> portest(monthintel, lags = seq(10, 40, 10), MonteCarlo = TRUE, 
+        ncores = ncores, squared.residuals = TRUE)
\end{Sinput}
\begin{Soutput}
 lags statistic     p-value
   10  34.67590 0.000999001
   20  52.04751 0.000999001
   30  64.65689 0.000999001
   40  72.26830 0.000999001
\end{Soutput}
\end{Schunk}


\subsection[Monte Carlo significance test for ARMA-GARCH models]{Monte Carlo significance test for \ARMA-\GARCH models}
\label{Example4}

While the \ARMA models specify the conditional mean of processes,
the \GARCH models specify the non-constant conditional variance of these processes.
Sometimes these two models are combined together to form the family of the \ARMA-\GARCH models.
A comprehensive account of these models is given
in the book \citep{Zivot2006} that also serves as the documentation
for the add-on modules of the statistical software \proglang{S-Plus}, \pkg{Finmetrics}.
Most of the functionality provided by \pkg{Finmetrics} for \GARCH
and related models involving \ARMA-\GARCH models are available with the 
\proglang{R} \pkg{fGarch} package \citep{Wuertz2013}.

As an illustrative example, we consider fitting the \ARMA-\GARCH model to the U.S. 
inflation \citep{Bollerslev1986}.
We used the quarterly GNP deflator from 1947-01-01 to 2010-04-01.
Data is available from the \pkg{fGarch} package as well as from our package 
with the name \code{GNPDEF}.
There are $n=254$ observations which we denoted by $z_t, t=1,\ldots, n$.
The inflation rate may be estimated by the logarithmic difference,
$r_t = \log(z_t)-\log(z_{t-1})$.

First, the function \code{garchFit()} from the \proglang{R} package \pkg{fGarch}
is used to fit the \ARMA(3,0)-\GARCH(1,1) model 

$r_t = 0.103 + 0.369 r_{t-1} + 0.223 r_{t-2} + 0.248 r_{t-3} + \epsilon_t$,
and
$\sigma_t^2 = 0.004 +  0.269 \epsilon_{t-1}^2 + 0.716 \sigma_{t-1}^2$.

Then the Monte Carlo \code{MahdiMcLeod} test is used to test the adequacy of this model.
As we mentioned in Example \ref{Example2}, the \code{portest()} function does not directly support
the fitted models with class \code{"fGARCH"} obtained by the \code{garchFit()}
function from the \pkg{fGarch} package.
For this reason, users must coded two functions as described before in
Section \ref{GoodnessOfFitTest} similar to the 
following code 

\begin{Schunk}
\begin{Sinput}
> # install.packages("fGarch") is needed
> require("fGarch")
> FitFGModel <- function(data){
+  GarchFit <- garchFit(formula = ~arma(3,0)+garch(1,1), data=data, trace=FALSE)     
+  GARCHOrder <- as.vector(GarchFit@fit$series$order)
+   res <- ts(residuals(GarchFit, standardize = TRUE))
+ 	p1 <- GARCHOrder[1]
+ 	q1 <- GARCHOrder[2]
+ 	p2<- GARCHOrder[3]
+ 	q2<- GARCHOrder[4]
+     n <- length(data)
+ 	mu <- as.vector(GarchFit@fit$par)[1]
+     ar <- as.vector(GarchFit@fit$par)[2:(p1+1)]
+     ma <- 0 
+    	omega <- as.vector(GarchFit@fit$par)[p1+q1+2]
+ 	alpha1 <- as.vector(GarchFit@fit$par)[(p1+q1+3):(p1+q1+p2+2)]
+ 	beta1 <- as.vector(GarchFit@fit$par)[(p1+q1+p2+3):(p1+q1+p2+q2+2)]
+     delta <- GarchFit@fit$params$delta
+ 	skew <- GarchFit@fit$params$skew
+ 	shape <- GarchFit@fit$params$shape
+ 	cond.dist <- GarchFit@fit$params$cond.dist
+ list(res=res,n=n,mu=mu,omega=omega,ar=ar,ma=ma,
+  alpha1=alpha1,beta1=beta1,delta=delta,skew=skew,shape=shape,cond.dist=cond.dist)
+ }
> SimFGModel <- function(model){
+    n <- model$n
+    mu <- model$mu
+    ar <- model$ar
+    ma <- model$ma
+    omega <- model$omega
+    alpha1 <- model$alpha1
+    beta1 <- model$beta1
+    delta <- model$delta
+    skew <- model$skew
+    shape <- model$shape
+    cond.dist <- model$cond.dist
+  spec <- fGarch::garchSpec(model=list(mu=mu,ar=ar,ma=ma,omega=omega,
+    alpha1=alpha1,beta1=beta1,delta=delta,skew=skew,shape=shape),cond.dist=cond.dist)
+   Sim.Data <- fGarch::garchSim(spec,n=n)
+  Sim.Data
+ }
> z <-ts(GNPDEF[,2], start=1947, freq=4)
> r <- 100*diff(log(z))
> FitfGarch <- FitFGModel(r)
> portest(FitfGarch,ncores=ncores,model=list(sim.model=SimFGModel,fit.model=FitFGModel),
+         pkg.name="fGarch",squared.residuals=FALSE)
\end{Sinput}
\begin{Soutput}
 lags statistic     p-value
    5  8.912642 0.001998002
   10 16.728762 0.002997003
   15 21.113277 0.003996004
   20 24.468216 0.010989011
   25 27.258753 0.022977023
   30 30.041855 0.035964036
\end{Soutput}
\end{Schunk}

The output from the Monte Carlo version of the portmanteau $\mathfrak{D}_m$ test statistic 
indicates that the fitted model is not an adequate model.

As we mentioned before in Example \ref{Example1}, one can always implement the 
portmanteau test statistics simply on the fitted residual.
In this case, no need to code the previous two functions, \code{FitFGModel()} and \code{SimFGModel()}, 
and the \proglang{R} code in such a case will be similar to the following one

\begin{Schunk}
\begin{Sinput}
>  Fit <- garchFit(formula = ~arma(3,0)+garch(1,1), data=r, trace=FALSE)     
>   res <- ts(residuals(Fit, standardize = TRUE))
>   portest(res,ncores = ncores,squared.residuals = FALSE)
\end{Sinput}
\begin{Soutput}
 lags statistic    p-value
    5  8.912642 0.05294705
   10 16.728762 0.02797203
   15 21.113277 0.03696304
   20 24.468216 0.05594406
   25 27.258753 0.07592408
   30 30.041855 0.10689311
\end{Soutput}
\begin{Sinput}
> detach(package:fGarch)
\end{Sinput}
\end{Schunk}

The results support that the fitted model is not an adequate model.  


\subsection[Monte Carlo significance test for threshold AR models]{Monte Carlo significance test for threshold \AR models}
\label{Example6}

The threshold autoregressive, \TAR, model developed by
\cite{Tong1978} and discussed in detail in \citet{Tong1990} 
provides a general flexible family for nonlinear time
series modeling that are useful in economics and many other applications.

The two-regime threshold autoregressive model (sometimes it referred to a two-regime
self-exciting threshold autoregressive, \SETAR(2), model) may be written 
\begin{eqnarray*}
  Z_t &=& \mu_{1}+\phi_{1,1} Z_{t-1} +\ldots+ \phi_{1,p_1} Z_{t-p_1} +\sigma_1 e_t, \mbox{ if } Z_{t-d}\leq r \\
  Z_t &=& \mu_{2}+\phi_{2,1} Z_{t-1} +\ldots+\phi_{2,p_2} Z_{t-p_2}+\sigma_2 e_t, \mbox{ if } Z_{t-d} > r, 
\end{eqnarray*}
where $\phi_{i,j}$, $j =1,\ldots,p_i~\&~i=1,2$ are the \AR coefficients,  
$e_t$ is the white-noise with mean zero and variance one, $\mu_{i}$ for $i=1,2$ are the means
of the two-regimes, $\sigma_{i}$ for $i=1,2$ are 
the standard deviation of disturbance terms, $r$ 
is the threshold, and $d$ is the delay parameter.

Further discussion of these and other topics involving \TAR models in \proglang{R}
are available in the textbook of \citet{CryerChan2008} and 
the \pkg{TSA} package introduced by \citet{ChanRipley2012}.

In this section, we use the stationary prey Didinium time series 
from the \code{veilleux} data frame available in the \pkg{TSA} package.
The prey Didinium time series dataset is available in this package with the 
name \code{prey.eq} has 57 numbers of prey individuals 
(Didinium natsutum, a protozoan) per/ml measured every 12 hours over a period of 35 days.
The experiment studied the population fluctuation of a prey-predator system; 
the prey is Paramecium aurelia, a unicellular ciliate protozon, whereas 
the predator species is Didinium natsutum.

\citet{CryerChan2008} and \citet{ChanRipley2012} fitted the threshold \AR model
using the logarithmic transformation to the number of predators,
where the order of the \AR lower and upper regimes are selected 
to be $p=1$ and $p=4$ respectively and the delay parameter $d=3$.

For diagnostic accuracy and adequacy of this model, we may 
use the Monte Carlo version of \code{MahdiMcLeod} test statistic as 
explained before in Example \ref{Example2}.
In this regard, we introduce 
the following two functions \code{fit.model()} and \code{sim.model()}.
As explained before, these two functions are needed for the Monte Carlo significance test 
in order to pass the argument \code{model} inside the \code{portest()} function.

\begin{Schunk}
\begin{Sinput}
> # install.packages("TSA") is needed
> require("TSA")
> FitModel <- function(data){
+     fit <- TSA::tar(y=log(data),p1=4,p2=4,d=3,a=0.1,b=0.9,print=FALSE)
+     res <- ts(fit$std.res)
+     parSpec <- list(res=res,fit=fit)
+   parSpec
+ }
> SimModel <- function(parSpec){
+     fit <- parSpec$fit
+   exp(tar.sim(fit)$y)
+ }
> data(prey.eq)
> portest(FitModel(prey.eq),ncores=ncores,model=list(SimModel,FitModel),pkg.name="TSA")
\end{Sinput}
\begin{Soutput}
 lags statistic   p-value
    5  2.588669 0.4295704
   10  4.125518 0.7332667
   15  5.792759 0.8631369
   20  7.421699 0.9160839
   25  9.195862 0.9530470
   30 10.633459 0.9760240
\end{Soutput}
\end{Schunk}

\begin{Schunk}
\begin{Sinput}
>  fit <- TSA::tar(y=log(prey.eq),p1=4,p2=4,d=3,a=0.1,b=0.9,print=FALSE)
>   res <- ts(fit$std.res)
>   portest(res,ncores=ncores)
\end{Sinput}
\begin{Soutput}
 lags statistic   p-value
    5  2.588669 0.6033966
   10  4.125518 0.8101898
   15  5.792759 0.8841159
   20  7.421699 0.9090909
   25  9.195862 0.9100899
   30 10.633459 0.9240759
\end{Soutput}
\begin{Sinput}
> detach(package:TSA)
\end{Sinput}
\end{Schunk}

One can also apply the Monte Carlo test on the fitted residual 
instead of applying this test on the fitted model (see the note in Example \ref{Example1}).
For both methods, the results suggest that the model is an adequate model.


\subsection[Monte Carlo significance test for seasonality]{Monte Carlo significance test for seasonality}
\label{Example7}

In this example we implement the portmanteau statistic on an econometric model of aggregate demand in the U.K.
to show the usefulness of using these statistics in testing the seasonality.
The data are quarterly, seasonally unadjusted in 1958 prices, covering the period
1957/3-1967/4 (with 7 series each with 42 observations), as published in Economic Trends 
and available from our package with the name \code{EconomicUK}. 
This data were disused by \citet{ProtheroWallis1976}, where 
they fit several models to each series and compared 
their performance with a multivariate model (See \citep[Tables 1-7]{ProtheroWallis1976}).

For simplicity, we select the first series, \code{Cn: Consumers' expenditure on durable goods},
and the first model $1a$ as fitted by \citet{ProtheroWallis1976} in Table 1.

\begin{Schunk}
\begin{Sinput}
> # install.packages("forecast") is needed
> require("forecast")
> cd <- EconomicUK[,1]
> cd.fit <- Arima(cd,order=c(0,1,0),seasonal=list(order=c(0,1,1),period=4))
\end{Sinput}
\end{Schunk}

After fitting \SARIMA$(0,1,0)(0,1,1)_4$, we apply the usual $\mathfrak{D}_m$ test statistic as well as the 
seasonal version of $\mathfrak{D}_m$ test statistic.
The asymptotic distribution and the Monte Carlo significance test suggest that the model is good.

\begin{Schunk}
\begin{Sinput}
> MahdiMcLeod(cd.fit,lags=c(5,10),season=1) ## Asympt. dist. for usual check
\end{Sinput}
\begin{Soutput}
 lags statistic       df   p-value
    5  1.700823 3.090909 0.6532001
   10  3.714068 6.857143 0.7999453
\end{Soutput}
\begin{Sinput}
> MahdiMcLeod(cd.fit,lags=c(5,10),season=4) ## Asympt. dist. for seasonal check
\end{Sinput}
\begin{Soutput}
 lags statistic       df   p-value
    5 0.6612291 3.090909 0.8918977
   10 1.5718612 6.857143 0.9771575
\end{Soutput}
\begin{Sinput}
> portest(cd.fit,lags=c(5,10),ncores=ncores)## MC check for seasonality
\end{Sinput}
\begin{Soutput}
 lags statistic   p-value
    5  1.700823 0.6013986
   10  3.714068 0.4795205
\end{Soutput}
\end{Schunk}


\subsection[Monte Carlo significance test for time series regression]{Monte Carlo significance test for time series regression}
\label{Example8}

We end this section by using the Monte Carlo significance 
test for time series regression application.
We consider the level of lake Huron from the year 1875 to 1972.
Data is available from package \pkg{datasets} with the name \code{LakeHuron}.
In this example, the level of the lake is regressed on the time 
after subtracting the value 1920 from each time value.
We fit the \ARIMAX(2,0,0) model using \code{Arima()} function from the 
contributed \proglang{R} package \pkg{forecast} \citep{HyndmanRazbash2015}
and implement the Monte Carlo 
version of \code{MahdiMcLeod} test statistic 
with 1000 replications and lags $1,2,\ldots,5$ for diagnosis the fitted model. 
The results indicate that the fitted model is an adequate one.

\begin{Schunk}
\begin{Sinput}
> require("forecast")
> fit.arima <- Arima(LakeHuron, order = c(2,0,0), xreg = time(LakeHuron)-1920)
> portest(fit.arima,lags=1:5,ncores = ncores)
\end{Sinput}
\begin{Soutput}
 lags  statistic   p-value
    1 0.03257799 0.5804196
    2 0.08741760 0.7242757
    3 0.11103807 0.8831169
    4 0.17518653 0.9290709
    5 0.28974267 0.9450549
\end{Soutput}
\end{Schunk}

As another illustrative example, we consider the annual \US macroeconomic data
from the year 1963 to 1982 with two variables, 
\code{consumption}: the real consumption and 
\code{gnp}: the gross national product.
Data was studied by \citet[Chapter 7, p. 221, Table 7.7]{Greene1993} and is available from 
the package \pkg{lmtest} \citep{Hothornpackage2015} under the name \code{USDistLag}.

First, we fit the distributed lag model as discussed in \citet[Example 7.8]{Greene1993} as follows,
\begin{eqnarray*}
\label{Greene1}
\hbox{\code{cons}} &\sim& \hbox{\code{gnp + cons1}}
\end{eqnarray*}

\begin{Schunk}
\begin{Sinput}
> # install.packages("lmtest") is needed
> require("lmtest")
> data("USDistLag")
> usdl <- stats::na.contiguous(cbind(USDistLag, lag(USDistLag, k = -1)))
> colnames(usdl) <- c("con", "gnp", "con1", "gnp1")
> fm1 <- lm(con ~ gnp + con1, data = usdl)
\end{Sinput}
\end{Schunk}

Then we write \proglang{R} code function \code{fn()} returns the 
generalized Durbin-Watson test statistic so that we can pass it to the 
argument \code{fn} inside the function \code{portest()}.
\begin{Schunk}
\begin{Sinput}
> fn <- function(obj,lags){
+      test.stat <- numeric(length(lags))
+        for (i in 1:length(lags))
+           test.stat[i] <- -sum(diff(obj,lag=lags[i])^2)/sum(obj^2)
+        test.stat
+ }
\end{Sinput}
\end{Schunk}

After that we apply the Monte Carlo version of the generalized Durbin-Watson test statistic
at lags 1, 2, and 3, using the nonparametric bootstrap residual, which
clearly detects a significant positive autocorrelation at lag 1. 

\begin{Schunk}
\begin{Sinput}
> portest(fm1, lags=1:3, test = "other", fn = fn, ncores = 4, innov.dist= "bootstrap") 
\end{Sinput}
\begin{Soutput}
 lags statistic    p-value
    1  1.356622 0.03196803
    2  2.245157 0.71128871
    3  2.488189 0.94205794
\end{Soutput}
\end{Schunk}

When residual autocorrelation is detected, sometimes simply 
taking first or second differences is all that 
is needed to remove the effect of autocorrelation \citep{McLeodYuMahdi2012}.

\begin{Schunk}
\begin{Sinput}
> fm2 <- lm(con ~ gnp + con1, data = diff(usdl,differences=1))
\end{Sinput}
\end{Schunk}

After differncing, the Monte Carlo version of the Durbin-Watson test statistic fail to reject the reject 
the null hypothesis of no autocorrelation and suggest that the differening model 
is an adequate one.

\begin{Schunk}
\begin{Sinput}
> portest(fm2, lags=1:3, test = "other", fn = fn, ncores = 4, innov.dist= "bootstrap")
\end{Sinput}
\begin{Soutput}
 lags statistic   p-value
    1  2.346099 0.7402597
    2  1.404779 0.2097902
    3  1.335600 0.2197802
\end{Soutput}
\begin{Sinput}
> detach(package:lmtest)
\end{Sinput}
\end{Schunk}


\section[Some Useful Functions]{Some useful functions}
\label{UsefulFunctions}

\subsection[Stationary and invertibility of ARMA/VARMA models]{Stationary and invertibility of \ARMA/\VARMA models}
\label{InvertQ}

The conditions for stationary and invertibility of the \VARMA$(p,q)$ process are the same as in
the pure \VAR$(p)$ and pure \VMA$(q)$ cases,
respectively \citep{Reinsel1997}.
The multivariate \VAR$(p)$ model of $k$-dimensional time series
$\bm{Z}_{t}=(Z_{1,t},\ldots,Z_{k,t})^{\prime}$ with mean
vector $\bm\mu$ and no deterministic equation is given by
\begin{equation}\label{VAR.Model1}
{\bm\Phi(B)Z}_{t}-\bm{\mu}=\bm{e}_{t},
\end{equation}
where
$\bm{\Phi(B)}=\bm {\mathbb{I}}_{k}-\bm{\Phi}_{1}\bm{B}-\cdots-\bm{\Phi}_{p}{\bm{B}}^{p}$,
can always expressed in the state-space $kp$-dimensional \VAR$(1)$ model in terms of
$\bm{Z^{\star}}_{t}=(\bm{Z}_{t}^{\prime},\ldots,\bm{Z}_{t-p+1}^{\prime})^{\prime}$
as
$\bm{Z^{\star}}_{t}=\bm\mu+\bm{\Phi^{\star}Z^{\star}}_{t-1}+\bm{e^{\star}}_{t},$
with
$\bm{e^{\star}}_{t}=(\bm{e}_{t}^{\prime},\bm{0}^{\prime},\ldots,\bm{0}^{\prime})^{\prime}$ and $\bm{\Phi^{\star}}$
equal to the $kp\times kp$ companion matrix associated with the \VAR$(p)$ operator $\bm{\Phi(B)},$ that is,
\begin{equation}\label{StationaryEquation}
\bm{\Phi^{\star}}=\left(%
\begin{array}{ccccc}
  \bm{\Phi}_{1} & \bm{\Phi}_{2} & \ldots & \ldots&\bm{\Phi}_{p} \\
  \bm {\mathbb{I}}_{k} & \bm{0} & \ldots &\ldots& \bm{0}\\
   \bm{0} & \bm {\mathbb{I}}_{k} & \bm{0}&\ldots&\bm{0} \\
  \vdots & \ddots & \ddots & \ddots& \vdots  \\
  \bm{0} & \bm{0} & \dots & \bm {\mathbb{I}}_{k} &\bm{0}\\
\end{array}%
\right)_{kp\times kp}.
\end{equation}
Similar to this, the $k$-dimensional \VMA$(q)$ process can be
represented in the state-space $kq$-dimensional \VMA$(1)$ model as
$\bm{Z^{\star}}_{t}=\bm\mu+\bm{e^{\star}}_{t}-\bm{\Theta^{\star}}\bm{e^{\star}}_{t-1},$
where $\bm{\Theta^{\star}}$ is defined by Equation~\ref{StationaryEquation} after replacing
$\bm{\Phi}_{i}$ by $\bm{\Theta}_{j}$ and $i=1,\ldots,p$ by $j=1,\ldots,q$,
$\bm{e^{\star}}_{t}$ by $\bm{Z^{\star}}_{t}$ and $\bm{Z^{\star}}_{t}$ by $\bm{e^{\star}}_{t}$.
In this light, the stationary condition that all roots of det$[\bm{\Phi(B)}]=0$
are greater than one in absolute value in the \VAR$(p)$
model is equivalent to the condition in the state-space \VAR$(1)$
representation that all eigenvalues of the $kp\times kp$
companion matrix $\bm{\Phi^{\star}}$ be less than one in absolute value,
and the invertibility condition that all roots of det$[\bm{\Theta(B)}]=0$
are greater than one in absolute value in the \VMA$(q)$ model is equivalent
to the condition in the state-space \VMA$(1)$ representation that all
eigenvalues of the $kq\times kq$ companion matrix $\bm{\Theta^{\star}}$ be less than one in absolute value.
The function \code{InvertQ()} in our package uses this techniques and
checks the validity of the stationary and invertibility
assumptions in the process.
The syntax of \code{InvertQ} is:

\begin{CodeInput}
InvertQ(coef).
\end{CodeInput}

It takes a numeric, matrix, or array of coefficients of \AR, \MA, \VAR,
or \VMA process as an argument and returns a warning message,
\code{"check stationary/invertibility condition !"}, only if
the process is not stationary or not invertible.

For checking stationarity of a univariate process with coefficients $\phi_1= 0.7$, 
$\phi_2=-0.3$ and $\phi_3=0.6$ type:

\begin{Schunk}
\begin{Sinput}
R> phi <- c(0.7,-0.3,0.6)
R> InvertQ(phi)
\end{Sinput}
\begin{Soutput}
Warning message:
In InvertQ(phi) : check stationary/invertibility condition !
\end{Soutput}
\end{Schunk}

The warning message in the output means that the stationary condition in the process is not valid.

For checking stationarity of a process with dimension $k=2$ from \VAR(2) with coefficients
$\bm{\Phi_1}=\left(%
\begin{array}{cc}
  0.5 & 0.1 \\
  0.4 & 0.5 \\
\end{array}%
\right)~\hbox{and}~
\bm{\Phi_2}=\left(%
\begin{array}{cc}
   0.0 & 0.0 \\
  0.3 & 0.0 \\
\end{array}%
\right),$ type:

\begin{Schunk}
\begin{Sinput}
> phi <- array(c(0.5, 0.4, 0.1, 0.5, 0, 0.3, 0, 0), dim = c(2, 2, 2))
> InvertQ(phi)
\end{Sinput}
\end{Schunk}
and for checking invertibility of a process with dimension $k=3$ from \VMA(1)
with coefficient
$\bm{\Theta}=\left(%
\begin{array}{ccc}
  0.5 & 0.0& 0.0\\
  0.1 & 0.1& 0.3\\
  0.0 & 0.2& 0.3\\
\end{array}%
\right),$ type:

\begin{Schunk}
\begin{Sinput}
> theta <- array(c(0.5, 0.1, 0, 0, 0.1, 0.2, 0, 0.3, 0.3), dim = c(3, 3, 1))
> InvertQ(theta)
\end{Sinput}
\end{Schunk}
while for checking stationarity and invertibility of process from \VARMA(1,1)
with coefficients
$\bm{\Phi}=\left(%
\begin{array}{cc}
  0.5 & 0.7 \\
  0.1 & 0.5 \\
\end{array}%
\right)~\hbox{and}~
\bm{\Theta}=\left(%
\begin{array}{cc}
   0.9 & 0.4 \\
  0.3 & 0.1 \\
\end{array}%
\right),$ type:

\begin{Schunk}
\begin{Sinput}
> phi <- array(c(0.5, 0.1, 0.7, 0.5), dim = c(2, 2, 1))
> InvertQ(phi)
\end{Sinput}
\end{Schunk}

\begin{Schunk}
\begin{Sinput}
R> theta <- array(c(0.9,0.3,0.4,0.1),dim=c(2,2,1))
R> InvertQ(theta)
\end{Sinput}
\begin{Soutput}
Warning message:
In InvertQ(theta) : check stationary/invertibility condition !
\end{Soutput}
\end{Schunk}

The warning message in the output means that the invertibility condition in the process is not valid.


\subsection[Simulation from seasonal/nonseasonal ARMA/VARMA models]{Simulation from seasonal/nonseasonal \ARIMA/\VARIMA models}
\label{Simulation}

We may use the function \code{varima.sim()} in the proposed
package to simulate seasonal/nonseasonal data from \VARIMA, \ARIMA process
described in Equations~\ref{SVARIMA.Model1} and \ref{SARIMA.Model1} respectively.
The simulated data may have a deterministic constant drift and
time trend term with non-zero mean.
The \code{varima.sim()} function is:

\begin{CodeInput}
varima.sim(model=list(ar = NULL, ma = NULL, d = NULL, ar.season = NULL, 
    ma.season = NULL, d.season=NULL,period=NULL), n, k = 1, constant = NA, 
    trend = NA, demean = NA, innov = NULL, 
    innov.dist = c("Gaussian", "t", "stable", "bootstrap"), ...),
\end{CodeInput}

where \code{ar} and \code{ma} are the univariate/multivariate autoregressive
and moving average parameters respectively, whereas
\code{ar.season} and \code{ma.season} are the univariate/multivariate seasonal 
autoregressive and seasonal moving average parameters respectively,
and should have class \code{"array"} or \code{"NULL"} in case of $k>1$.
The arguments \code{constant} and \code{trend} represent
coefficients of the deterministic equation
and the argument \code{demean} stands for the mean of the series.
For $k=1$, these parameters can be entered with
class \code{"numeric"}, \code{"array"}, or \code{"NULL"}.
$\bm{d}=(d_1,\ldots,d_k),~d_i\geq 0$ is the usual differencing order,
whereas $\bm{d.season}=(ds_1,\ldots,ds_k),~ds_i\geq 0$ is the seasonal 
differencing at seasonal order $s$.
The argument \code{innov.dist} specifies the distribution that will be used 
to generate the innovation series.
The default distribution is Gaussian, but users may use the $t$
{\footnote{For $t$-distribution, users need to select the degrees of freedom as 
an integer number passes the argument \code{dft}}} or the stable distribution 
or the nonparametric bootstrap method. 
The argument \code{innov} is an optional argument that can used 
to include an initial univariate or
multivariate innovations series associated with the stable distribution 
or the nonparametric bootstrap method.
To simulate time series from models with infinite variance innovations,
then the argument \code{innov.dist = "stable"} in the function 
\code{varima.sim()} must be selected. 
In this case, optionally users may include an initial innovation series via the 
argument \code{innov} to estimate the four stable parameters, 
\code{Alpha, Beta, Scale}, and \code{Location}, that are needed for generating data 
with an infinite variance innovation. 
Otherwise, for given values of the argument \code{par.stable},
the function \code{varima.sim()} calls the function \code{rStable()}
to generate innovations from stable distributions, 
$S(\alpha,\beta,\gamma,\delta)$ corresponds to \code{Alpha, Beta, Scale, Location} respectively.
The syntax of \code{rStable()} in our package is defined as follows:

\begin{CodeInput}
rStable(n, Alpha, Beta, Scale = NULL, Location = NULL).
\end{CodeInput}

The parameters \code{Alpha} represent the index parameter of the stable
distribution and must be entered as a numeric or vector
with values in the range $(0,2]$.
The skewness parameters \code{Beta} values must be in the range $[-1, 1]$.
The scale parameters, and the location parameters, \code{Scale, Location}, can have \code{NULL}
values or real values in the range $(-\infty,\infty)$.

Our function \code{rStable()} may be considered an extension to
the function \code{rStable()} in the \proglang{R} \pkg{fBasics} package
\citep{Wuertz2014} as it can be used for generating univariate
or multivariate data from independent stable distributions.

The function \code{varima.sim()} checks the conditions of
stationary or invertibility or both in the simulated process by calling the
function \code{InvertQ()}, then it determines recursively the impulse response coefficients,
$\psi_{l}$ or $\Psi_{l}$, $l=1,2,\ldots$, by solving the
equation $\psi(B)=\phi^{\#{-1}}(B)\theta^{\#}(B)$ or $\bm{\Psi(B)}=\bm{\Phi^{\#{-1}}(B)\Theta^{\#}(B)}$
given in Equations \ref{SVARIMA.Model2} and \ref{SARIMA.Model2} respectively \citep{Reinsel1997}.
Finally it represents this process in terms of an infinite \MA or \VMA filter, as follows
\begin{eqnarray}
\label{ImpulseResponseVMA}
Z_{t}&=&\mu+e_t+\psi_1e_{t-1}+\psi_2e_{t-2}+\cdots\\
\label{ImpulseResponseVMA2}
\bm{Z}_{t}&=&\bm\mu+\bm{e}_{t}+\bm{\Psi}_{1}\bm{e}_{t-1}+\bm{\Psi}_{2}\bm{e}_{t-2}+\cdots
\end{eqnarray}
The impulse response coefficients $\psi_{l}$ or $\Psi_{l}$ are
calculated from the function \code{ImpulseVMA()} in our package at sufficiently
large number.
That is, we set a large value to the argument \code{trunc.lag}
in the functions \code{ImpulseVMA()} or \code{varima.sim()} at which the models in
Equation~\ref{ImpulseResponseVMA} or Equation~\ref{ImpulseResponseVMA2} need to be truncated,
so that the weight of the impulse response coefficient at this truncated value is negligible.
If \code{trunc.lag} is not given in the function \code{varima.sim()} then the
infinite \MA or \VMA representation will be truncated at the
minimum value of 100 or least integer value of $n/3$, where $n$ is
the length of the simulated series.
If \code{trunc.lag} is \code{NULL} is used in the
function \code{ImpulseVMA()} then the truncation value will be $p+q$, where
$p$ and $q$ represent the order of autoregressive and moving average respectively.

More information and examples about the \code{varima.sim()} function can be found
on the \pkg{portes} manual documentation.

\subsubsection[Simulation example 1]{Simulation example 1}
\label{SimulationExample1}

To generate a univariate white noise series with length 200, simply type:
\begin{Schunk}
\begin{Sinput}
> z1 <- varima.sim(n=200)
\end{Sinput}
\end{Schunk}
and to generate a bivariate white noise series with length 200, type:
\begin{Schunk}
\begin{Sinput}
> z11 <- varima.sim(n=200, k = 2)
\end{Sinput}
\end{Schunk}
whereas to generate a multivariate white noise series with dimension $200\times 4$, type:
\begin{Schunk}
\begin{Sinput}
> z111 <- varima.sim(n=200, k = 4)
\end{Sinput}
\end{Schunk}


\subsubsection[Simulation example 2]{Simulation example 2}
\label{SimulationExample2}

To generate a univariate time series with length 100 from
\AR(2,1) with drift equation $ 2 + 0.01 t$,
mean $\mu=0,$ \ARMA coefficients $\phi_1=0.7,\phi_2=0.2, \theta=-0.5$
and t-distribution of 5 degrees of freedom, type:

\begin{Schunk}
\begin{Sinput}
> n <- 100
> phi <- array(c(0.7, 0.2), dim = c(1, 1, 2))
> theta <- -0.5
> z2 <- varima.sim(list(ar = phi, ma = theta), n, constant = 2, trend = 0.01, 
+       innov.dist = "t", dft = 5, trunc.lag = 50)
\end{Sinput}
\end{Schunk}

In this example the infinite \MA series is truncated at the value 50,
however users can use any desirable truncation value.
If the \code{trunc.lag} is not given, then the 
default \code{trunc.lag = min$(100, n/3)$}, where $n$ 
is the length of the series, will be implemented.


\subsubsection[Simulation example 3]{Simulation example 3}
\label{SimulationExample3}

In this example, we simulate two univariate seasonal series, each of a size 300, from 
a \SARIMA$(2,1,0)\times(0,1,1)_s$ model, where $s$ is the seasonal period $s=12$ for the first 
series and $s=4$ for the second one. 
For both series, we generate data using Gaussian innovations, where 
$\phi_1=1.3$ ,$\phi_2=-0.35$, and $\Theta_s=0.8$.
In this example, the default value \code{trunc.lag = min(100, 300/3) = 100} 
is used to truncate the simulated infinite series. 

\begin{Schunk}
\begin{Sinput}
> n <- 300
> set.seed(12754)
> phi <- c(1.3, -0.35)
> theta.season <- 0.8
> z3<-varima.sim(list(ar=phi,d=1,ma.season=theta.season,d.season=1),n=n)
> z33<-varima.sim(list(ar=phi,d=1,ma.season=theta.season,d.season=1,period=4),n=n)
\end{Sinput}
\end{Schunk}

\begin{figure}[!ht]
\begin{center}
\includegraphics{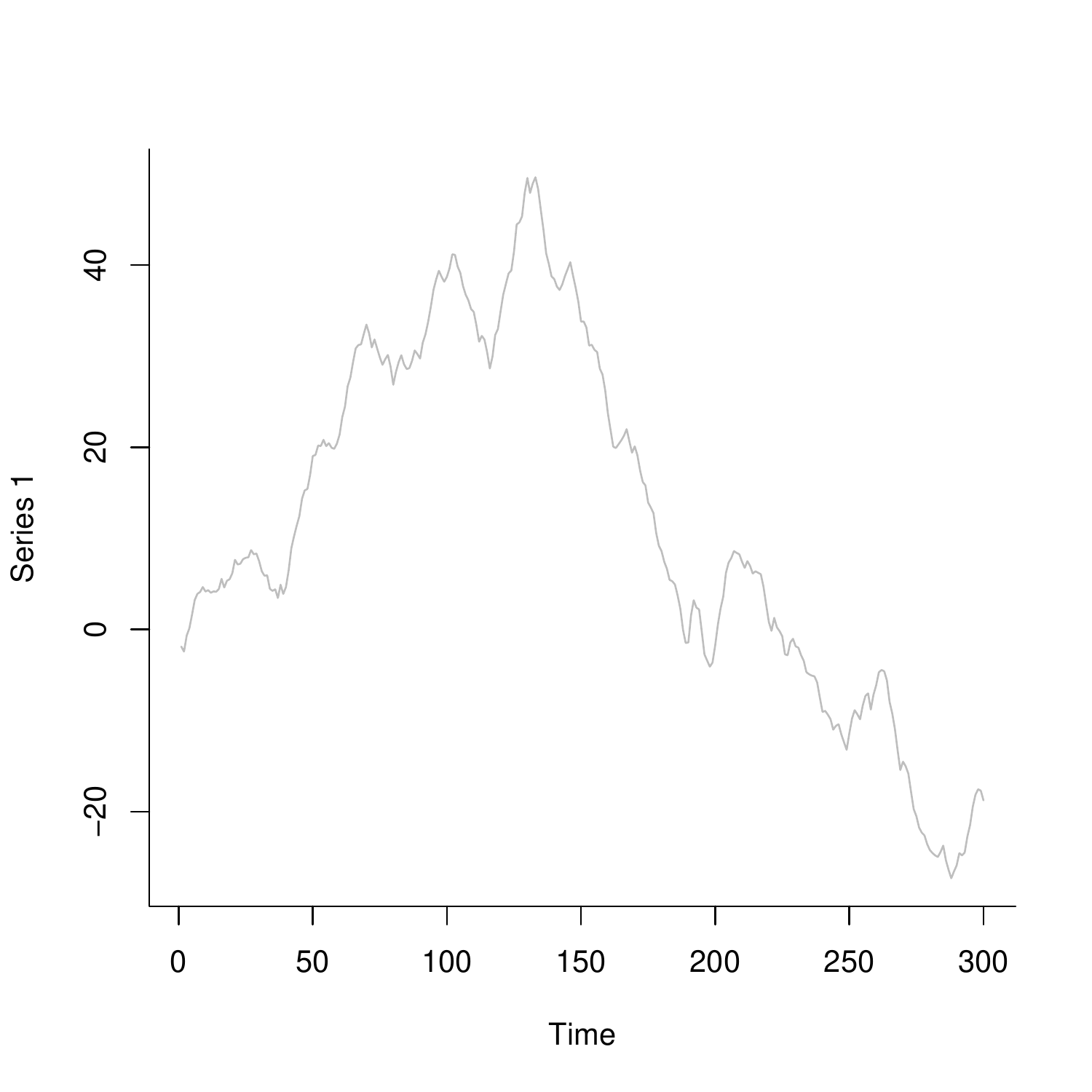}
\end{center}
\caption{A simulated time series of example~\ref{SimulationExample3}.}
\label{fig1}
\end{figure}


\subsubsection[Simulation example 4]{Simulation example 4}
\label{SimulationExample4}

In this example we simulate a bivariate time series of
length 200 from a \VARMA(1,1) model with mean
$\mu = (2,5)^{\prime}$, drift equation $\bm{a}+\bm{b}t=(1,4)^{\prime}+(0,0.04)^{\prime}t$,
and coefficient matrices
$\bm{\Phi}=\left(%
\begin{array}{cc}
  0.5 & 0.1 \\
  0.4 & 0.5 \\
\end{array}%
\right),~\hbox{and}~
\bm{\Theta}=\left(%
\begin{array}{cc}
   0.5 & 0.6 \\
  -0.7 & 0.3 \\
\end{array}%
\right),$
where
$\bm{a}_{t}$ are generated from multivariate normal distribution
with mean vector zero and covariance matrix
$\bm{\Gamma}_{0}=\left(%
\begin{array}{cc}
  1.00 & 0.71 \\
  0.71 & 1.00 \\
\end{array}%
\right)$,
where the default value \code{trunc.lag = ceiling(200/3) = 67} 
is used to truncate the simulated infinite series.

\begin{Schunk}
\begin{Sinput}
> set.seed(123)
> n <- 200
> phi <- array(c(0.5, 0.4, 0.1, 0.5), dim = c(2, 2, 1))
> theta <- array(c(0.5, -0.7, 0.6, 0.3), dim = c(2, 2, 1))
> sigma <- matrix(c(1, 0.71, 0.71, 1), 2, 2)
> constant <- c(1, 4)
> trend <- c(0, 0.04)
> demean <- c(2, 5)
> z4 <- varima.sim(list(ar = phi, ma = theta), n = n, k = 2,
+    sigma = sigma, constant = constant, trend = trend, demean = demean)
\end{Sinput}
\end{Schunk}

\begin{figure}[!ht]
\begin{center}
\includegraphics{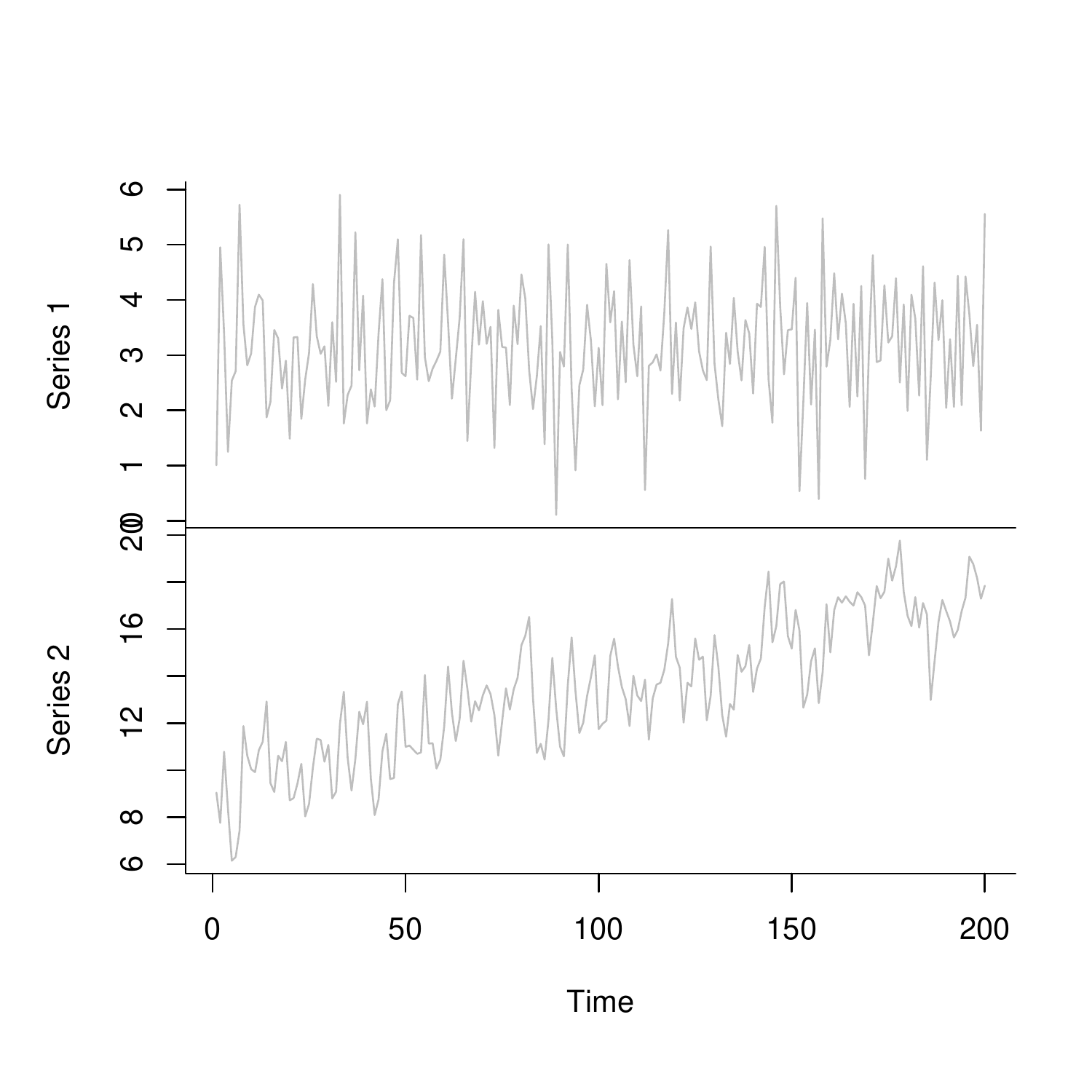}
\end{center}
\caption{A simulated time series of example~\ref{SimulationExample4}.}
\label{fig2}
\end{figure}


\subsubsection[Simulation example 5]{Simulation example 5}
\label{SimulationExample5}

In this example, we simulate a trivariate seasonal vector series,  
\SVARIMA$(1,0,0)\times(1,0,0)_{12}$, of dimension $300\times 3$. 
The parameters are 
$\bm{\Phi}=\left(%
\begin{array}{ccc}
  0.5 & 0.5 & 0.0 \\
  0.4 & 0.0 & 0.0 \\
  0.1 & 0.3 & 0.1 \\
\end{array}%
\right)$
and 
$\bm{\Phi^{\bullet}_{12}}=\left(%
\begin{array}{ccc}
  0.00 & 0.50 & 0.00 \\
  0.25 & 0.10 & 0.25 \\
  0.00 & 0.40 & 0.60 \\
\end{array}%
\right)$.

The process have mean c(10, 0, 12), drift equation $a + b \times t$, 
where $a$ and $b$ are vectors equal $(2,1,5)$ and $(0.01,0.06,0)$ respectively.
The innovations series are generated from multivariate normal distribution,
where the default value \code{trunc.lag=ceiling(100/3)=34} is used to  
truncate the simulated infinite series.

\begin{Schunk}
\begin{Sinput}
> set.seed(1234)
> k <- 3
> n <- 300
> phi <-  array(c(0.5,0.4,0.1,0.5,0,0.3,0,0,0.1),dim=c(k,k,1))
> phi.season <-  array(c(0,0.25,0,0.5,0.1,0.4,0,0.25,0.6),dim=c(k,k,1))
> constant <- c(2,1,5)
> trend <- c(0.01,0.06,0)
> demean <- c(10,0,12)
> z5 <- varima.sim(list(ar=phi,ar.season=phi.season),n=n,k=k,constant=constant,
+   trend=trend,demean=demean)
\end{Sinput}
\end{Schunk}
\begin{figure}[!ht]
\begin{center}
\includegraphics{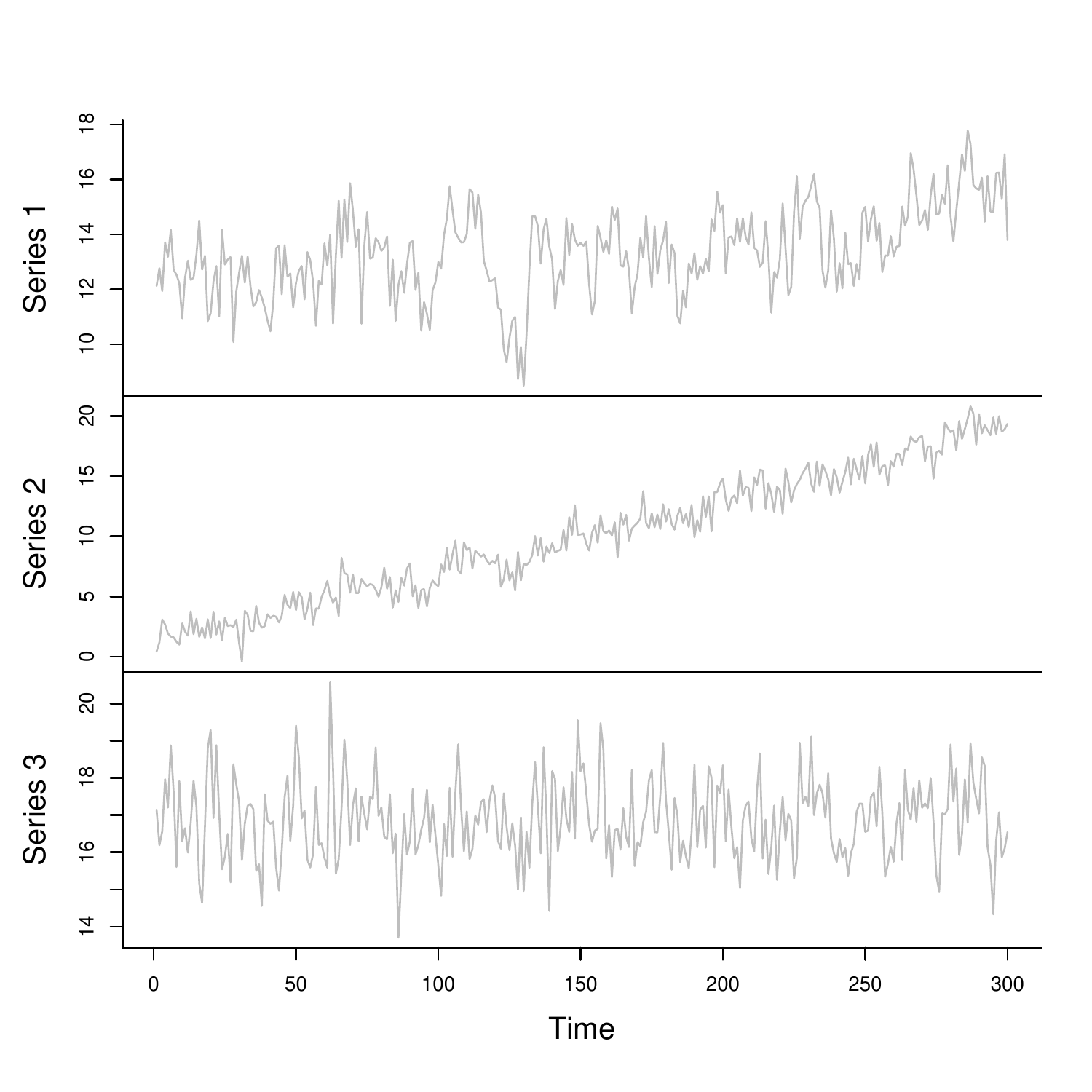}
\end{center}
\caption{A simulated time series of example~\ref{SimulationExample5}.}
\label{fig3}
\end{figure}


\subsubsection[Simulation example 6]{Simulation example 6}
\label{SimulationExample6}

In this example, we simulate a bivariate white noise series from a multivariate 
$t4$-distribution, then we the nonparametric bootstrap method to generate a seasonal 
\SVARIMA of order $(0,d,0)\times(0,0,1)_{12}$ with $d = (1,0)^{\prime}, n= 250, k = 2$, and 
$\bm{\Theta^{\bullet}_{12}}=\left(%
\begin{array}{cc}
   0.5 & 0.1 \\
   0.4 & 0.3 \\
\end{array}%
\right)$.

\begin{Schunk}
\begin{Sinput}
> set.seed(1234)
> z6 <- varima.sim(n=250,k=2,innov.dist="t",dft=4)
> theta.season=array(c(0.5,0.4,0.1,0.3),dim=c(2,2,1)) 
> z66 <- varima.sim(list(ma.season=theta.season,d=c(1,0)),n=250,k=2,
+                  innov=z6,innov.dist="bootstrap")
\end{Sinput}
\end{Schunk}
\begin{figure}[!ht]
\begin{center}
\includegraphics{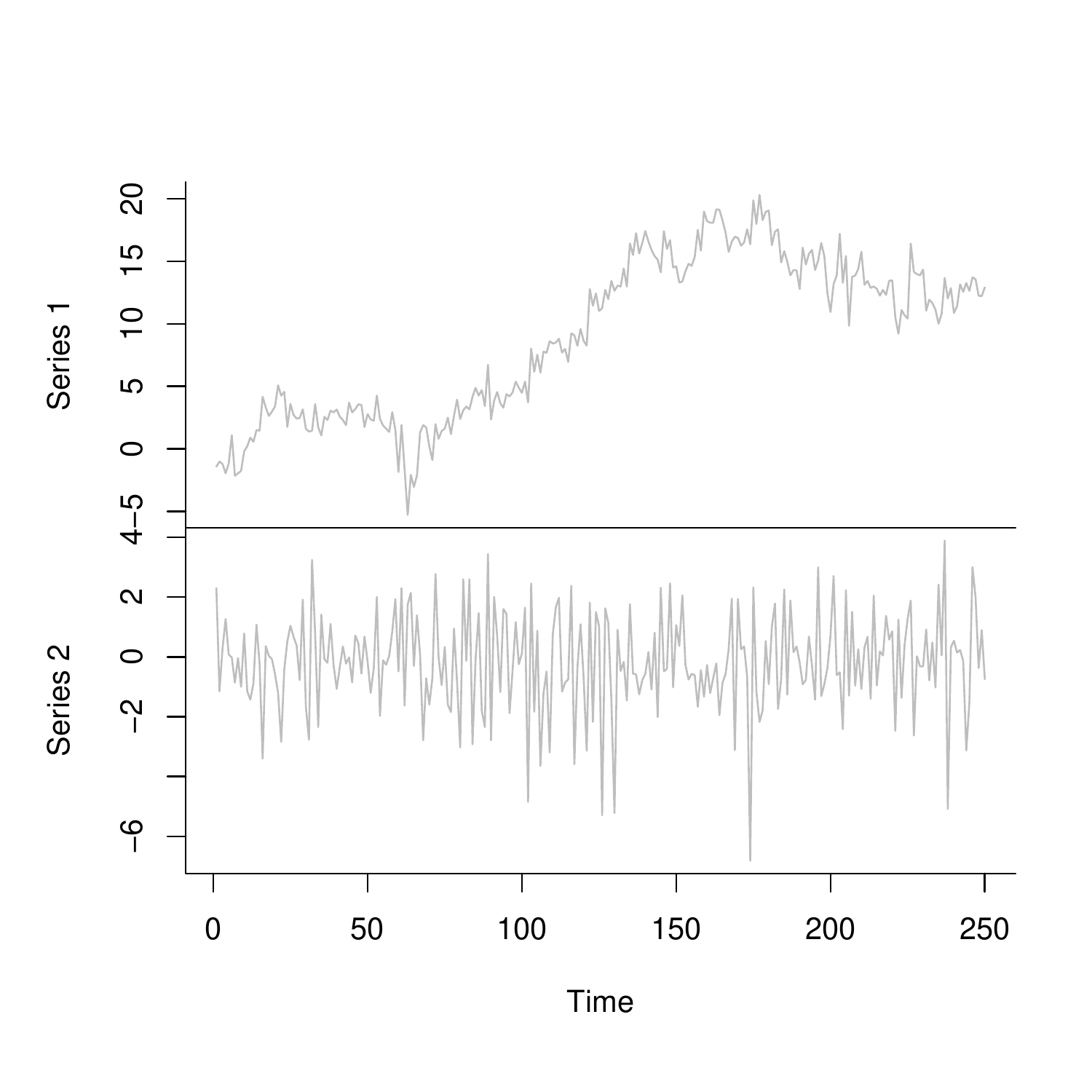}
\end{center}
\caption{A simulated time series of example~\ref{SimulationExample6}.}
\label{fig4}
\end{figure}


\subsubsection[Simulation example 7]{Simulation example 7}
\label{SimulationExample7}

Finally, to simulate an \VAR(2) process of length 600 with coefficient matrices
$\bm{\Phi_1}=\left(%
\begin{array}{cc}
  0.8 & 0 \\
  0.0 & -2 \\
\end{array}%
\right),$
$\bm{\Phi_2}=\left(%
\begin{array}{cc}
   -0.5 & 0.0 \\
    0.0 & 0.0 \\
\end{array}%
\right),$
covariance matrix
$\bm{\Gamma}_{0}=\left(%
\begin{array}{cc}
  1.0 & 0.5 \\
  0.5 & 1.0 \\
\end{array}%
\right),$
and error term from a stable distribution with
$\alpha = (1.3,1.6)^{\prime},~\beta = (0,0.2)^{\prime},~\gamma = (1,1)^{\prime},~\delta = (0,0.2)^{\prime}$, type:

\begin{Schunk}
\begin{Sinput}
> set.seed(1234)
> k <- 2
> n <- 600
> phi <- array(c(0.8,0,-2,0),dim=c(k,k,1))
> theta <- array(c(-0.5,0,0,0),dim=c(k,k,1))
> sigma <- matrix(c(1,0.5,0.5,1),k,k)
> Alpha <- c(1.3,1.6)
> Beta <- c(0,0.2)
> Scale <-c(1,1)
> Location <-c(0,0.2)
> par.stable <- c(Alpha, Beta, Scale , Location)
> z7 <- varima.sim(list(ar = phi, ma = theta), n = n, k = k,
+    sigma = sigma, innov.dist = "stable", par.stable = par.stable)
\end{Sinput}
\end{Schunk}
\begin{figure}[!ht]
\begin{center}
\includegraphics{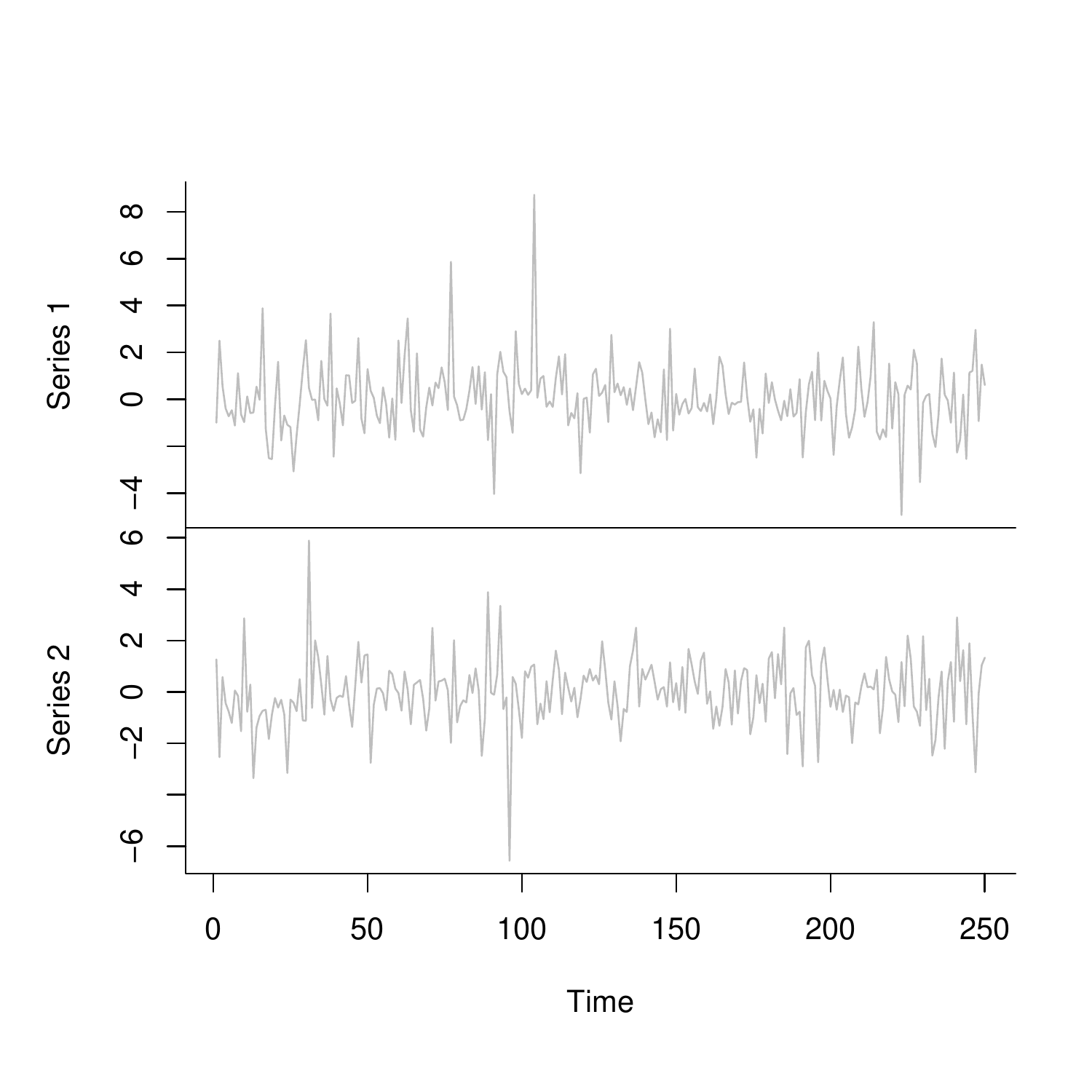}
\end{center}
\caption{A simulated time series of example~\ref{SimulationExample7}.}
\label{fig5}
\end{figure}
In this example the series is truncated at default value: \code{trunc.lag} = min$(100,600/3)=100$.

\subsection[Fit parameters to stable distribution]{Fit parameters to stable distribution}
\label{fitstable}

The quantile estimation method of \citet{McCulloch1986} is implemented in
the \proglang{R} function \code{fitstable()}.
This method is highly reliable, fast and reasonably efficient especially
bearing in mind that in most applications there is a lot of data.
The function \code{fitstable()} is:

\begin{CodeInput}
fitstable(x),
\end{CodeInput}
where \code{x} represents the data frame.
This function is used under the assumption of that the variables
of the vector \code{x} are independents.
The output of this function is a $k$ rows represents the number of
variables in the vector x, and 4 columns with named components
\code{alpha, beta, scale}, and \code{location} associated to
the stable parameters, \code{Alpha, Beta, Scale}, and \code{Location} respectively.

In the following example we use the function \code{fitstable()} 
to estimate the stable parameters
for a simulated data from \ARMA(2,1) model with errors from stable distribution.

\begin{Schunk}
\begin{Sinput}
> set.seed(54368)
> n <- 1000
> phi <- c(1.3, -0.35)
> theta <- 0.1
> Alpha<- 1.5
> Beta <- 0
> Scale <- 1
> Location <- 0
> par.stable <- c(Alpha, Beta, Scale, Location)
> x<-varima.sim(list(ar=phi,ma=theta),n,innov.dist="stable",par.stable=par.stable)
> fitstable(x)
\end{Sinput}
\begin{Soutput}
    Alpha       Beta    Scale Location
 1.436127 -0.2743106 6.330361 5.219509
\end{Soutput}
\end{Schunk}

As another example we use the function \code{rStable()} to simulate
a bivariate independent data of size $100$ from a stable distribution
with heavy tail then we use the function \code{fitstable()}
to estimate the stable parameters as follows

\begin{Schunk}
\begin{Sinput}
> set.seed(54368)
> Alpha <- c(1.3, 1.6)
> Beta <- c(0, 0.2)
> Scale <- c(1, 1)
> Location <- c(0, 0.2)
> sim.series1 <- rStable(100, Alpha, Beta, Scale, Location)
> fitstable(sim.series1)
\end{Sinput}
\begin{Soutput}
    Alpha      Beta     Scale   Location
 1.550165 -0.276085 0.9314066 0.08272251
 1.705539  1.000000 0.9158591 0.57164890
\end{Soutput}
\end{Schunk}

Another illustration of using the function \code{varima.sim()},
note that we can use this simulated series, \code{sim.series1}, as an initial
innovation series in order to simulate a new series from a stable distribution as follows

\begin{Schunk}
\begin{Sinput}
> set.seed(123)
> sim.series2<-varima.sim(n=100,k=2,innov=sim.series1,innov.dist="stable")
> fitstable(sim.series2)
\end{Sinput}
\begin{Soutput}
    Alpha       Beta     Scale   Location
 1.443405 0.01110013 0.9458293 -0.1171760
 1.533737 0.91503986 0.8772945  0.7865066
\end{Soutput}
\end{Schunk}


\section{Conclusion}
\label{conclusion}

Our package \pkg{portes} version 4.0 is available from the Comprehensive
\proglang{R} Archive Network, \proglang{CRAN},
at \url{http://CRAN.R-project.org/package=portes} used by many
high quality publishes refereed papers such as
\citep{GallagherFisher2015,CuiFisherWu2014,FisherGallagher2012,MahdiMcLeod2012}.
It implements the Monte Carlo test in a very convenient way, especially if we are
running \proglang{R} in batch mode using the \pkg{parallel} 
package with multicore computers.
We believe that many \proglang{R} users may find that the \pkg{portes}
package is convenient for simulating time series
from seasona/nonseasonal \SARIMA and more generally from \SVARIMA
models with or without deterministic equation
where innovations have finite or infinite variances. 
This package is also useful for estimating parameters from stable distributions.




\bibliographystyle{jss}
\bibliography{portest}

\end{document}